\documentclass[aps,prx,reprint]{revtex4-1}
\usepackage{amsmath}
\usepackage{amssymb}
\usepackage{amsfonts}
\usepackage{graphicx}
\usepackage{cancel}
\renewcommand{\rm}[1]{\mathrm{#1}}

\begin{document}

\title{Tunable-Cavity QED with Phase Qubits}
\author{J. D. Whittaker}
\author{F.~C.~S.~da~Silva, M.~S.~Allman, F.~Lecocq,
K.~Cicak, A.~J.~Sirois, J.~D.~Teufel, J.~Aumentado}
\author{R.~W.~Simmonds}
\email{simmonds@boulder.nist.gov} \affiliation{National Institute
of Standards and Technology, 325 Broadway St, Boulder, Colorado
80305, USA}

\begin{abstract}
We describe a tunable-cavity QED architecture with an \mbox{rf SQUID}
phase qubit inductively coupled to a single-mode, resonant cavity
with a tunable frequency that allows for both microwave readout of tunneling and dispersive measurements of the qubit. Dispersive measurement is well characterized by a three-level model, strongly dependent on qubit
anharmonicity, qubit-cavity coupling and detuning. A tunable cavity frequency provides a way to strongly vary both the qubit-cavity detuning and coupling strength, which can reduce Purcell losses, cavity-induced dephasing of the qubit, and residual bus coupling for a system with multiple qubits. With our qubit-cavity system, we show that dynamic control over the cavity frequency enables one to avoid Purcell losses during coherent qubit evolutions and optimize state readout during qubit measurements. The maximum qubit decay time $T_1 = 1.5\,\mu$s is found to be limited by surface dielectric losses from a design geometry similar to planar transmon qubits.
\end{abstract}

\maketitle

\section{INTRODUCTION}
\label{Intro}
Quantum information processors and simulators made from many
superconducting qubits and cavities seem feasible in the near
future, as two, three, and four-qubit processors have already
demonstrated rudimentary algorithms and error correction \cite{YaleTwoQubitAlgorithm,
YaleThreeQubitErrorCorr,MartinisFourQubitAlgorithm,Devoret2013}. There has also been a revolution in improving superconducting qubit coherence times by both placing qubits within a 3D cavity \cite{3DTransmon,Rigetti2012} and through geometrical and materials improvements on chip \cite{DevoretGeometry,IBM,Patel2013,Pappas}. Both of these strategies have reduced dielectric losses \cite{Dielec} and provided nearly two orders-of-magnitude improvement in coherence times. Also, the advent of quantum-limited amplifiers has allowed single-shot, quantum non-demolition (QND) readout of qubits with fidelities above 90\%, which has lead to key demonstrations of quantum feedback, heralded state initialization, and teleportation \cite{Vijay2012,Johnson2012,Riste2012,Campagne2013,Steffen2013}.
Even with these improvements, there is still a need to find compact circuit architectures that can manage a limited spectrum of possible microwave frequencies, offering the possibility of
increasing the total number of qubits and cavities in a larger system. Ongoing refinements to fast, repeatable nondestructive QND measurements of qubits and ancillas will also be crucial to enable the routine performance of high fidelity quantum feedback, teleportation, and error correction \cite{Devoret2013}. Also, a full system architecture must, at the same time, avoid enhanced qubit energy decay \cite{YaleMultiPurcell}, dephasing \cite{Bertet2005,BoissonneaultDephase, Sears, Slichter}, and stray qubit-qubit coupling within a bus-like structure \cite{Devoret2013,CavityBus,StateTran}.

\subsection{Cavity QED with Transmon Qubits}

Transmon qubits have become widespread in circuit-based cavity quantum electrodynamics (QED) architectures with long coherence times and many key demonstrations \cite{Transmon, CavityBus,
YaleTwoQubitAlgorithm, YaleThreeQubitErrorCorr, IBM}. State
discrimination has been accessible through the state-dependent
dispersive shift of a coupled cavity's resonance frequency
\cite{Grajcar2004,Transmon}, well suited to performing QND measurements, with
various modifications that allow for single-shot readout
\cite{DispJBA, SingleShotQED, ReedOut}, joint-state readout
\cite{JointDisp}, the observation of quantum jumps
\cite{Qjump}, and the stabilization of Rabi oscillations \cite{Vijay2012}, along with most of the key demonstrations mentioned earlier. Although there has been dramatic progress, there are still some drawbacks related to the cavity. It can be a source for spontaneous emission (via the Purcell effect \cite{YaleMultiPurcell}),
dephasing of the qubit energy levels due to photon number
fluctuations \cite{Bertet2005,BoissonneaultDephase, Sears, Slichter}, and
residual coupling between multiple qubits in a cavity-bus
architecture \cite{StateTran, CavityBus}. Although some approaches, in principle, can avoid these issues by using a static cavity with tunable coupling between either the qubit and the cavity \cite{Allman2010,HouckTCoup2} or the cavity and its feedline \cite{EyobSete}, they have not yet been thoroughly tested. And while the development of a
``Purcell filter'' \cite{PurcellFilter} has been employed to reduce
spontaneous emission centered at a fixed qubit frequency over a relatively narrow bandwidth, it cannot eliminate the other drawbacks mentioned above. 

\subsection{rf~SQUID Phase Qubits}

For almost a decade, \mbox{rf SQUID} phase qubits \cite{TLSDecoherence}
have made steady progress, leading to remarkably successful
multi-qubit-cavity systems \cite{Coupled, StateTran,
MartinisFourQubitAlgorithm}. However, even with a clear
understanding of dielectric loss mechanisms \cite{Dielec}, long
coherence times have been lacking, with all energy relaxation times
$<1\,\mu$s, apart from one unique device with a crystalline silicon capacitor \cite{Patel2013}. Furthermore, phase qubits have relied on tunneling
events for state discrimination, which destroys the qubit, creates quasiparticles, and emits broadband microwave radiation crosstalk that can spoil the state of other coupled qubits or cavities \cite{Coupled,
StateTran, CrosstalkTheory, CavityCrosstalk}. Although simultaneous qubit measurement
\cite{Coupled} has been sufficient for key demonstrations, many of
the pitfalls discussed above are difficult to avoid when more than
one simultaneous measurement is required. Ultimately, the macroscopic quantum tunneling measurement technique is too harmful for cavity QED systems and probably too destructive to enable practical error correction \cite{Devoret2013}. Hence, there is a strong motivation to develop a dispersive measurement strategy for rf~SQUID phase qubits.

As we will see, moving to a tunable-cavity QED architecture is a natural choice
for performing dispersive measurements of \mbox{rf SQUID} phase qubits. Phase qubits can have large changes in their anharmonicity $\alpha$ that can strongly reduce the size of the state-dependent dispersive shifts $\chi$, especially for larger qubit-cavity detunings $\Delta_{01}$. A tunable cavity's frequency $\omega_c$ can be adjusted to compensate for any reductions in $\alpha$, by decreasing $|\Delta_{01}|$. With inductive coupling, as discussed later, it is also possible to take advantage of a tunable qubit-cavity coupling strength $g$ as well. Simultaneously increasing $g$, while decreasing $|\Delta_{01}|$ by only controlling $\omega_c$ through a flux bias $\phi_c$, significantly increases the dispersive shifts, $\chi$. Ideally, the strategy would be to simply place the qubit frequency sufficiently far below the maximum cavity frequency during qubit operations to maintain qubit coherence. When strong dispersive qubit measurements are required, the cavity frequency $\omega_c$ would be sufficiently lowered towards the qubit's operation frequency $\omega_{01}$, by applying a flux pulse through $\phi_c$. In this way, it is possible to maintain or increase the size of the dispersive shifts $\chi$, but only over short time periods during the measurement mode of operation. This should achieve the best of both worlds: long coherence times with strong, non-destructive QND measurements.

Surprisingly, not much attention has been given to the widely
tunable nonlinearity of the phase qubit \cite{Shalibo2012,Shalibo2013}. At their maximum frequency and
flux insensitive ``sweet-spot'', \mbox{rf SQUID} phase qubits can look
nearly harmonic ($|\alpha_R|\approx\,0.1\%$). At their lowest
frequencies (where they are typically metastable), the
anharmonicity can grow by nearly two orders-of-magnitude to
transmon levels ($|\alpha_R|\approx\,10\%$). This makes the \mbox{rf SQUID} phase qubit a `multi-purpose' quantum circuit element, as it can be tuned
appropriately to behave as a harmonic oscillator \cite{Osborn2007,Shalibo2012}, qudit \cite{Neeley2009}, or a qubit \cite{TLSDecoherence}. Recently, nonlinearity introduced into a high-Q 3D cavity has been used to generate complex photon states of microwave light \cite{Kirchmair2013}. Significant improvements in phase qubit coherence (as discussed later) could allow for rapid, on-demand creation of complex photon states \cite{Shalibo2013}. Phase qubits can also explore other types of rich physical behavior. For example, simple circuit modifications could lead to more interesting energy level structures for artificial atoms, with a Hamiltonian that is widely tunable. These features, along with the ease of coupling inductively or capacitively, are desirable for developing circuit
architectures for quantum simulators \cite{Buluta2009}. Additionally, phase qubits can also explore macroscopic quantum tunnneling phenomenon \cite{Schwartz1985,Devoret1985}, whose investigation could be enhanced through continuous dispersive measurements. 

\subsection{Tunable-Cavity QED}

In this work, we first introduce the concept of a tunable-cavity QED architecture that provides a way to perform dispersive measurements with dynamic control over the
qubit-cavity detuning and coupling strength. This offers a number of
key improvements over the use of a fixed frequency readout cavity. By
dynamically increasing the qubit-cavity detuning and decreasing the
qubit-cavity coupling, it should be possible to
significantly reduce both energy relaxation
\cite{YaleMultiPurcell} and qubit dephasing
\cite{BoissonneaultDephase, Sears, Slichter}, and, if part of a
bus-architecture, minimize residual qubit-qubit coupling
\cite{StateTran, CavityBus}. Independent control over the cavity
frequency also eliminates the need to change the qubit frequency,
allowing the qubit to stay fixed at any optimal value. This
approach also relaxes certain design constraints imposed by
limited (or variable) qubit nonlinearity or a fixed cavity
frequency, allowing for decreases in qubit-cavity coupling or
increases in cavity-feedline coupling ($\kappa/2\pi>20$~MHz, in this work). For example, the latter enables a faster cavity response, an increased number of photons delivered during measurement, and an improved signal to noise ratio, reducing the readout averaging time \cite{Gambetta2008,PurcellFilter}.
Other benefits include rapid removal of qubit dephasing cavity photons following a strong measurement and enhanced ground state thermalization \cite{PurcellFilter}.

Next, we describe an experimental implementation of a tunable-cavity QED architecture capable of (destructive and non-destructive) single-shot readout \cite{SingleShotQED, RFReadout, MartinisMultiplexed}. Unlike one similar experiment with a transmon qubit coupled capacitively to a tunable, multi-mode coplanar waveguide cavity \cite{Sandberg2009}, we employ a lumped-element cavity with inductive coupling to an \mbox{rf SQUID} phase qubit. This approach takes full advantage of a simple, flux tunable single-mode cavity resonance and a widely tunable qubit-cavity coupling strength. This architecture can be readily extended to other systems (i.e., transmons, flux qubits, quantum dots).   

We characterize both the tunable cavity and the phase qubit with spectroscopic measurements, extracting many of the system's circuit parameters. The sensitivity of the cavity frequency with flux provides a convenient way to perform rapid microwave readout of traditional tunneling measurements \cite{RFReadout,MartinisMultiplexed} of the phase qubit. The ability to perform single-shot tunneling measurements is helpful for rapidly characterizing phase qubits both spectroscopically and in the time-domain with Rabi and Ramsey oscillations. This allows us to quickly extract the qubit's anharmonicity $\alpha$, the energy relaxation time $T_1$, and the (inhomogenous-broadened) dephasing time $T_2^{(*)}$ as a function of qubit frequency. We then test the tunable-cavity QED approach by performing static dispersive measurements for multiple qubit and cavity frequencies with large variations in qubit anharmonicity, qubit-cavity coupling, and qubit-cavity detuning. We verify that the size of the full dispersive shifts $2\chi$ depends on all three of these factors and is well characterized by models that describe a three-level artificial atom coupled to a cavity. 

We then explore energy losses across the entire phase qubit spectrum for multiple cavity frequencies. We verify, for static operation with dispersive readout, that by strategically placing the cavity frequency at an optimal value, this architecture allows the qubit to avoid energy loss from the Purcell effect, with $T_1$ values that are mostly limited by dielectric losses and coupling to flux bias feedlines. This has not been achieved with the Purcell filter \cite{PurcellFilter}, which has only protected the qubit from Purcell loss at one filter frequency over a relatively narrow bandwidth. In addition, by utilizing the fast flux control of the cavity's frequency, we dynamically avoid the Purcell effect during coherent phase qubit evolution by rapidly increasing the qubit-cavity detuning to isolate the qubit and then dynamically change the cavity frequency for qubit measurements, optimizing the cavity flux for tunneling readout. Finally, we show that single-layer, planar transmon-like construction of rf~SQUID phase qubits using simple fabrication techniques gives relaxation times $>1\mu$s. This tunable cavity QED architecture has helped to fully characterize the loss experienced by \mbox{rf SQUID} phase qubits in the absence of strongly coupled dc SQUIDs \cite{RFReadout}, typically used for tunneling readout \cite{TLSDecoherence, Fast, Neeley2008}. 

\section{THE TUNABLE-CAVITY QED CONCEPT}
\label{TCQED}
In circuit-QED, the coupling of a superconducting qubit to a resonant cavity is generally described by the Jaynes-Cummings
hamiltonian:

\begin{equation}\label{eq:H}
H = H_q
+ H_c + \hbar
g\left(a^\dag\sigma^{-}+a\sigma^{+}\right)
\end{equation}
where $H_q = \hbar\omega_{01}\left(\sigma_z/2\right)$, $H_c = \hbar\omega_c\left(a^\dag a+1/2\right)$ describe the uncoupled qubit and cavity, $\omega_{01}$
and $\omega_c$ are the qubit and cavity transition frequencies,
$\sigma_z$ is the qubit state operator (with eigenvalues $\pm 1$),
$a^\dag$ ($a$) represents the raising (lowering) operator for the
cavity, and $\sigma^{+}$ ($\sigma^{-}$) represents the same for
the qubit. When the ``bare'' frequencies of the qubit and cavity match ($\omega_{01}=\omega_c$), the energy levels of the two systems hybridize, leading to an avoided-level crossing or normal-mode splitting of size $2g$ at this resonance frequency.

\begin{figure}[t!]
  \includegraphics[width=3.3in]{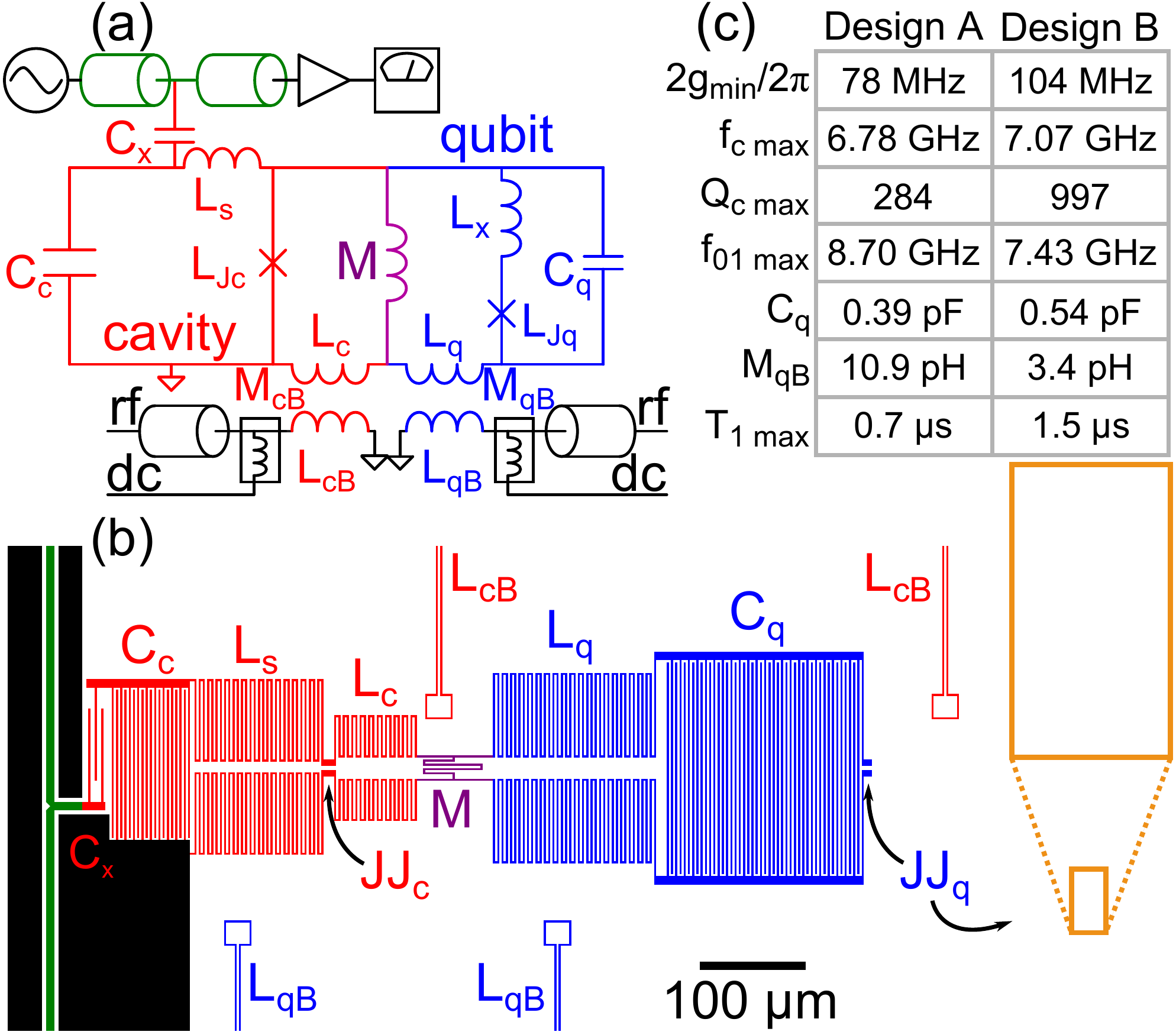}\\
  \caption{
(Color online) (a) Circuit schematic showing the cavity (red)
and qubit (blue) sub-circuits, the $Z_o=50\,\Omega$ microwave
feedline (green), their shared inductance $M$ (purple), and their
respective gradiometric bias lines $L_{cB}$ (red) and $L_{qB}$
(blue). (b) Fabrication layout for design $B$ and SEM-image of angle-evaporated Josephson junction. (c) Table highlighting differences between designs $A$ and
$B$. Values common to both designs are $L_s=1.79$~nH, $\mathcal{L}_c=L_c+M=0.75$~nH, $\mathcal{L}_q=L_q+M=2.5$~nH, $L_x=0.27$~nH, $I_{oc}=I_{oq}\approx 0.32\,\mu$A, and
$M=61$~pH. Both the phase qubit and the cavity were based on
\mbox{rf SQUID} designs with ratios
$\mathcal{L}_c/L_{Jc} = 0.7$ ($\mathcal{L}_q/L_{Jq}=2.8$)  for the cavity (qubit) (see text).}
 \label{Fig1}
\end{figure}

Qubit measurements are performed by driving the
microwave cavity at or near its resonance frequency and monitoring
the response for qubit state-dependent dispersive shifts. This
technique relies on the interaction strength $g$ between the qubit and
the cavity, as described by Eq.~(\ref{eq:H}). When the qubit frequency is far detuned from the cavity, $\Delta_{01} = \omega_{01}-\omega_c$ and
$|\Delta_{01}|\gg g$, the system's energy
level structure is slightly modified from the uncoupled case. A
unitary transformation and expansion of the Hamiltonian leads to a
frequency shift $\chi=(g^2/\Delta_{01})$ for the cavity that
depends on the qubit state \cite{Transmon},
$\omega_c\rightarrow\omega_c + \chi\sigma_z$. This result is
strictly valid only for two-level systems. For multi-level qubits,
with more transitions $\omega_{ab} = \left(E_b-E_a\right)/\hbar$,
the anharmonicity $\alpha = \omega_{12}-\omega_{01}$ due to the
next higher level of the qubit plays an important role in
determining the full dispersive shift $2\chi$. A three-level model \cite{Transmon, Boissonneault, ThreeLevelQubit} predicts a
dispersive shift of,
\begin{equation}\label{eq:chi}
\chi=
\frac{\left(g^2/\Delta_{01}\right)}{\left(1+\Delta_{01}/\alpha\right)}
\end{equation}
For both \mbox{rf SQUID} phase \cite{TLSDecoherence} and transmon
\cite{Transmon} qubits, the relative anharmonicity $\alpha_R =
\alpha/\omega_{01}$ is small ($|\alpha_R|\lesssim$\,\%10), so that
staying in the dispersive limit requires
$\Delta_{01}\gtrsim\alpha$, generally leading to significantly
smaller dispersive shifts compared to the two-level system result
(except in the ``straddling regime'' \cite{Transmon}, not
considered here \cite{Boissonneault2}).

For our tunable-cavity QED system, the coupling is provided by a mutually shared inductor, $M$ (see Fig.~\ref{Fig1}). The shared energy $\hbar g$ due to $M$ is then $M I_q I_c$, where $I_q$ ($I_c$) is the current flowing through $L_q$ ($L_c$). By looking at the schematic in Fig.~\ref{Fig1}(a), it is apparent that the size of these currents must depend on the value of the Josephson inductance within each \mbox{rf SQUID} loop. Specifically, as the Josephson inductance $L_{Jq}$ or $L_{Jc}$ increases, then the size of $I_q$ or $I_c$ must also increase. This situation then leads to a coupling strength $g$ that increases as the cavity
frequency $\omega_c$ decreases. If we neglect any small contributions from $L_x$ (since $L_x/L_{Jq}\ll 1$) and the self capacitance of the Josephson junctions (since $C_J/C_x\ll 1$ for $x=q,c$), we find that when $g\ll\omega_c\approx\omega_q$,
\begin{equation}\label{eq:g}
g=g_0\left[\left(\frac{\omega_{c0}}{\omega_c}\right)- \left(\frac{\omega_{c0}}{\omega_s^2}\right)\omega_c\right]
\end{equation}
with $g_0 = \omega_{q0}M/2\sqrt{\mathcal{L}_q \mathcal{L}_c}$, $\omega_s=1/\sqrt{L_s C_c}$, $\omega_{x0}=1/\sqrt{\mathcal{L}_x C_x}$ for $x = q,c$. Notice that the maximum (minimum) coupling rate is defined by the minimum (maximum) cavity frequency and that the coupling is never ``off'', $g\neq 0$. The addition of a separate tunable coupling element \cite{Allman2010} could provide $g=0$, but requires an additional independent flux control line. 

\begin{figure}[!t]
  \includegraphics[width=3.3in]{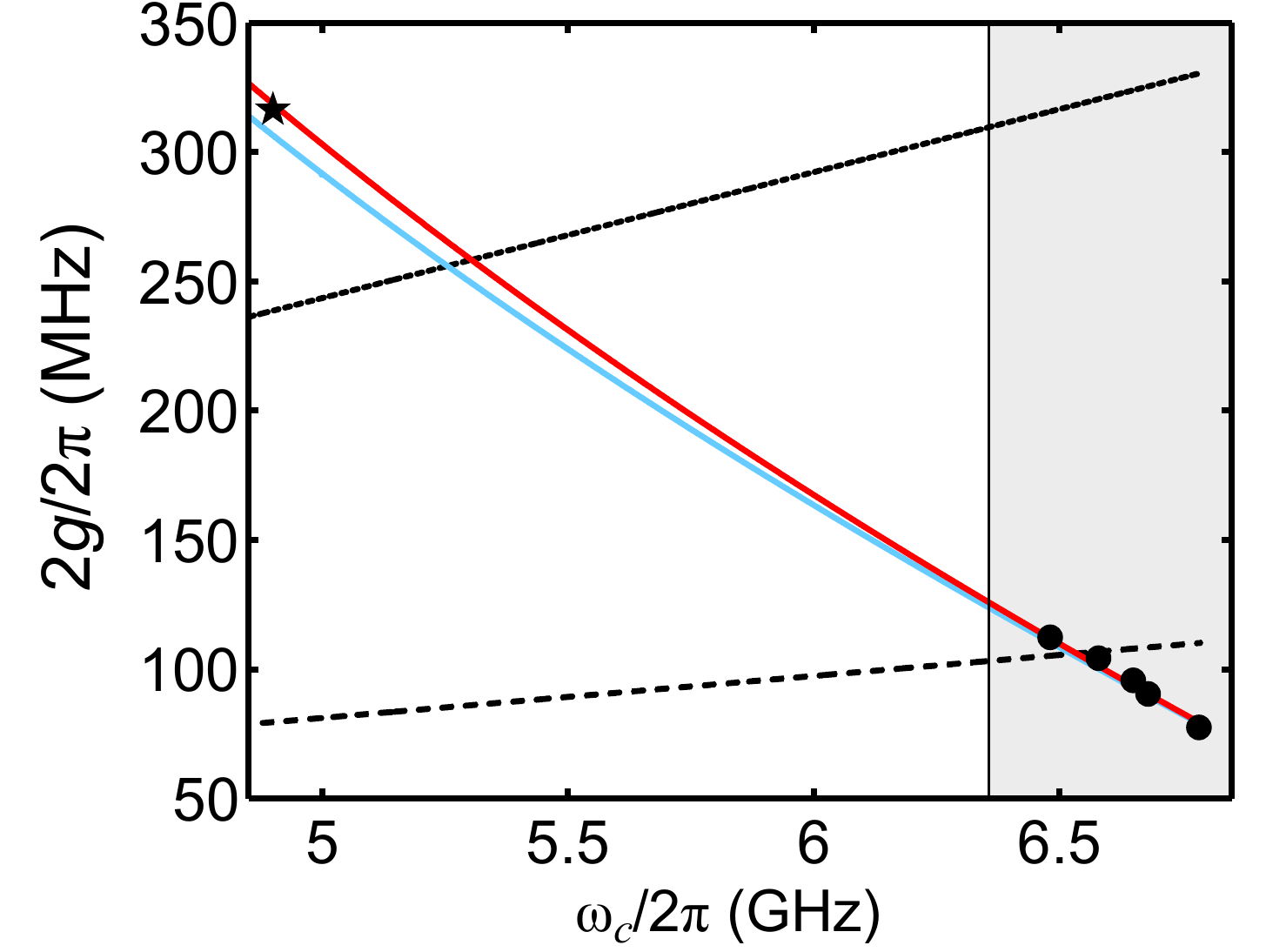}\\
  \caption{
(Color online) Coupling rate $2g/2\pi$ (design $A$) as a function of cavity frequency $\omega_c/2\pi$. The solid red (blue) line is the prediction from Eq.~(\ref{eq:g}) (including $L_x$ and $C_J$'s). The \mbox{(dotted)} dashed line is the prediction for capacitive coupling with $C=15$~fF ($C=5$~fF). The solid circles were measured spectroscopically (see text). At lowest cavity frequency, the solid {\Large $\star$} results from a fit to the Purcell data, discussed later in section~\ref{TCQEDC}. The gray region highlights where the phase qubit (design $A$) remains stable enough for operation (see text).
}\label{Fig2}
\end{figure}

An example of coupling rate as a function of cavity frequency is shown for design $A$ in Fig.~\ref{Fig2}. The experimental points were extracted directly from the vacuum Rabi splittings found in the spectroscopic measurements of both the cavity and the qubit, as discussed later in section~\ref{TCQEDA} (see Fig.~\ref{Fig7}). The minimum qubit frequency we could accurately measure spectroscopic splittings was about 6.48~GHz, limited by the onset of macroscopic quantum tunneling of the ground state (see gray region in Fig.~\ref{Fig2}). At the lowest cavity frequency (denoted by {\Large $\star$}), the coupling rate $2g/2\pi$ results from a fit to the Purcell curves discussed later in section~\ref{TCQEDC}. Measuring Purcell loss as a function of qubit frequency offers an alternative way of extracting the coupling strength between the qubit and the cavity, even when the two systems are far detuned, precluding one's ability to accurately capture a vacuum Rabi splitting or resolve very small dispersive shifts.

Tunable coupling is a direct consequence of combining a static coupling element and a frequency tunable cavity. The particular relationship between $g$ and $\omega_c$ depends on the details of the circuit design. For comparison, in Fig.~\ref{Fig2} we also show the prediction for the coupling rate had we used a single coupling capacitor $C=5$~fF or $C=15$~fF connecting the two capacitors $C_q$ and $C_c$, while removing $M$ and replacing $L_q\rightarrow L_q+M$ and $L_c\rightarrow L_c+M$ to roughly maintain the same frequency range. In this case, the coupling strength is linear and increases with increasing cavity frequency, $g = \omega_c C/2\sqrt{C_q C_c}$, and is never ``off'', $g\neq 0$. Notice that capacitive coupling is well-suited for a tunable-cavity QED architecture with transmon qubits, however the dynamic range for changing the coupling rate is far weaker than the inductive case. 

The coupling method we choose depends on how we wish to operate the device and on what qubit measurement strategy we wish to use. It is important to note, that in any case, we only need to control one tunable parameter, the cavity frequency $\omega_c$, in order to change both the coupling strength $g$ and the detuning $\Delta_{01}$ in some optimal way. By choosing the relative frequencies of the qubit and cavity appropriately, the goal is to satisfy two simple criteria: (1) the coupling strength $g$ and detuning $|\Delta_{01}|$ should vary in opposition to each other in order to avoid unwanted cavity interactions, and (2) $\omega_c$ should be set to a value that optimizes the separation between the two cavity frequencies for adjacent flux states in the qubit's rf SQUID loop when using tunneling measurements (see section~\ref{QBA}), \emph{or} when performing dispersive qubit readout (see section~\ref{TCQEDB}) as the nonlinearity of the qubit decreases, $g$ should increase and $|\Delta_{01}|$ should decrease in order to maintain sufficiently large dispersive shifts $2\chi$ for improved detection efficiency. 

For the inductively coupled phase qubit, if the cavity is operated at a frequency far above much of the qubit's operational spectrum, then we can satisfy both of these criteria. With the cavity at its maximum frequency, $|\Delta_{01}|$ can be made large, while the coupling $g$ is at its minimum and relatively small. This condition is the ``coherent mode of operation'' that isolates the qubit from the cavity, minimizing Purcell loss, dephasing, and possible stray bus-coupling. In order to optimize qubit readout, the cavity frequency $\omega_c$ must be adjusted. Ideally, this would be done dynamically during the ``measurement mode of operation''. Here, $\omega_c$ is rapidly shifted by applying a fast flux bias pulse to $\phi_c$. When using tunneling measurements, a fast flux pulse is sent to $\phi_q$ to make the measurement, and then the cavity is shifted to an optimal frequency for microwave readout, as discussed later in section~\ref{QBA}. Performing strong dispersive measurements requires one to shift the cavity frequency $\omega_c$ to lower values, increasing $g$ while decreasing $|\Delta_{01}|$, to increase the full dispersive shift $2\chi$. Even for larger phase qubit frequencies, where the nonlinearity $\alpha$ is reduced, the cavity frequency can be optimally lowered in order to increase $g$ and reduce $|\Delta_{01}|$ (according to Eq.~\ref{eq:chi}), so that $2\chi$ can be increased. During the ``measurement mode of operation'', Purcell loss, dephasing, and possible stray bus-coupling can increase significantly, so shifts in $\omega_c$ should only occur for a short time, long enough to readout the qubit state. Notice that dynamic operation is required if one wishes to optimize \emph{both} qubit isolation (performance) and the quality of the qubit measurement.

Fast qubit measurements can be accomplished in this architecture with a phase qubit using tunneling or dispersive readout. As mentioned previously, when using tunneling measurements, one should still optimize the cavity position, increasing the signal-to-noise ratio (SNR) to enable single-shot readout. For fast dispersive readout, several factors must be optimized. First, the optimal SNR is achieved when the dispersive shift is increased to $2\chi=\kappa$, with SNR$_\mathrm{max}=2n\kappa T_1\eta$, and $\eta\approx 1/(N+1)$ is the detection efficiency \cite{Gambetta2008}. Here, the noise added by the amplification chain, usually dominated by the first stage amplifier with sufficient gain, should be ideally quantum-limited, adding a minimum number $N$ of noise photons. Next, the response rate of the cavity $\kappa$ should be large and dominated by feedline (output) coupling, not internal losses, in order to determine the rate of outgoing itinerant photons carrying qubit information. The bandwidth $B$ of the amplifier chain must be sufficiently wide for this information to pass through rapidly, basically $B\geq\kappa$. Quantum-limited amplification with parametric amplifiers with bandwidths of $B/2\pi=20$~MHz have already been used to see quantum jumps \cite{Qjump} and larger bandwidths have also been helpful for deterministic teleportation experiments \cite{Steffen2013}. Notice that the amplitude of the microwave tone, which puts on average $n$ photons in the cavity, also determines SNR$_\mathrm{max}$, but is limited by the critical photon number \cite{Transmon,Gambetta2008,BoissonneaultDephase} $n_{\mathrm{crit}}=(\Delta_{01}/2g)^2$. This maintains the dispersive approximation and the QND character of the readout. Finally, the qubit state should have sufficiently long $T_1$, so that its rate of energy decay is much smaller than the rate information is gained by the cavity about the qubit's state, or $1/T_1\ll\kappa$. Thus, its important to balance increases in $2\chi$ to satisfy the optimization condition $2\chi=\kappa$, while not reducing $T_1$ too severely from the Purcell effect. 

\begin{figure*}[!t]
  \centering
  \includegraphics[width=6in]{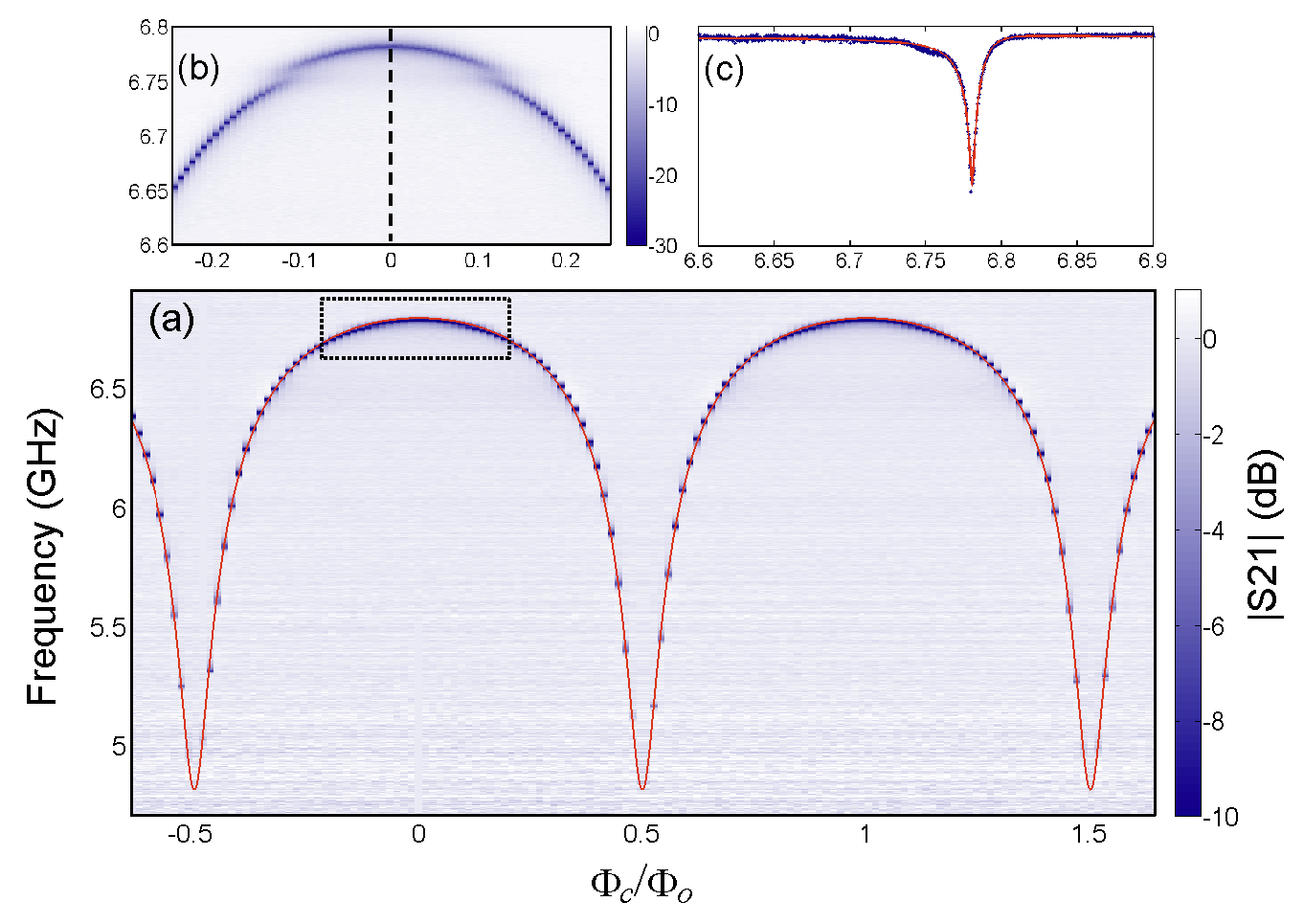}
  \caption{
(Color online) (a) Cavity spectroscopy (design $A$) while sweeping the cavity flux bias with the qubit far detuned, biased at its maximum frequency. The solid line is a fit to the model including the junction capacitance. (b) Zoom-in near the maximum cavity frequency
showing a slot-mode. (c) Line-cut on resonance along the dashed
line in (b) with a fit to a skewed Lorentzian (solid line).}
 \label{Fig3}
\end{figure*}

\begin{figure*}[!t]
  \centering
  \includegraphics[width=6in]{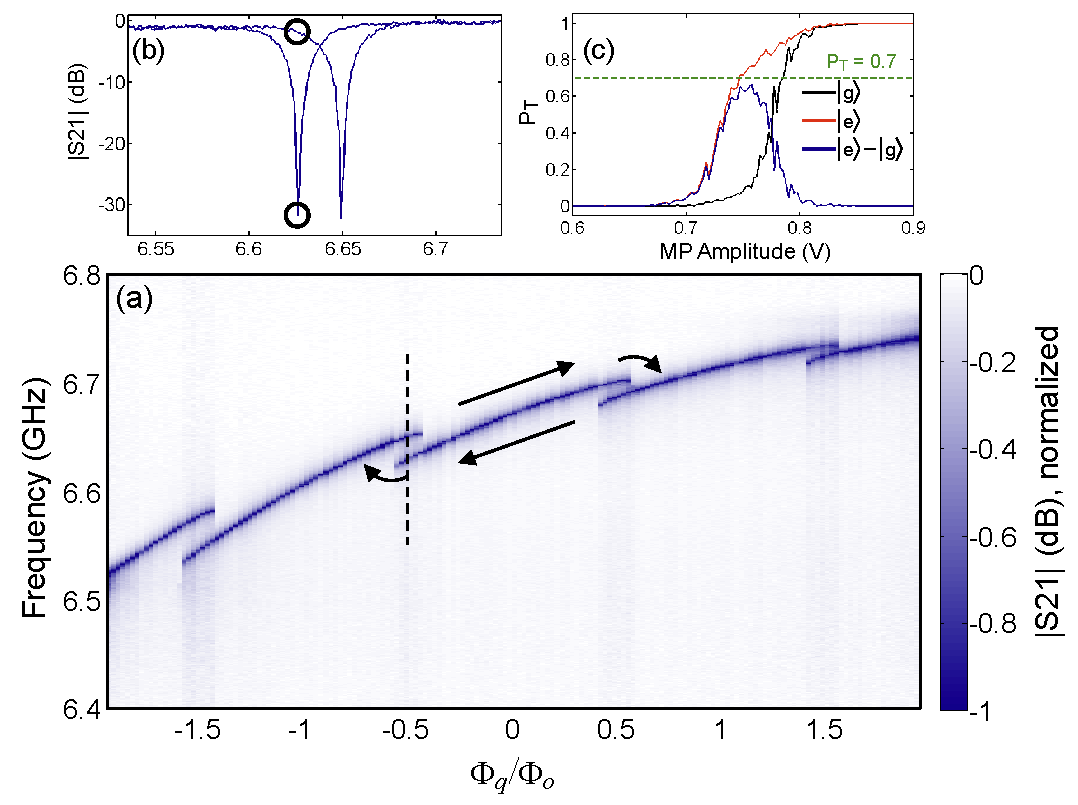}
  \caption{
(Color online) (a) The cavity response (design $A$) as a the qubit flux bias is swept. Two different data sets (with the qubit reset at
$\phi_q=\pm2$) have been overlayed to show the double-valued or
hysteretic regions. The straight arrows show the direction of
operation and the curved arrows indicate tunneling transitions.
(b) Cavity transmission for each qubit flux state near
$\phi_q=-0.5$, dashed line in (a). (c) Tunneling probability $P_T$
for the $|g\rangle$-state and $|e\rangle$-state versus measure
pulse (MP) amplitude. The maximum raw measurement contrast is
obtained by maximizing the difference between the $P_T$-curves,
$|e\rangle-|g\rangle$.}
 \label{Fig4}
\end{figure*}

\begin{figure*}[!t]
  \centering
  \includegraphics[width=6in]{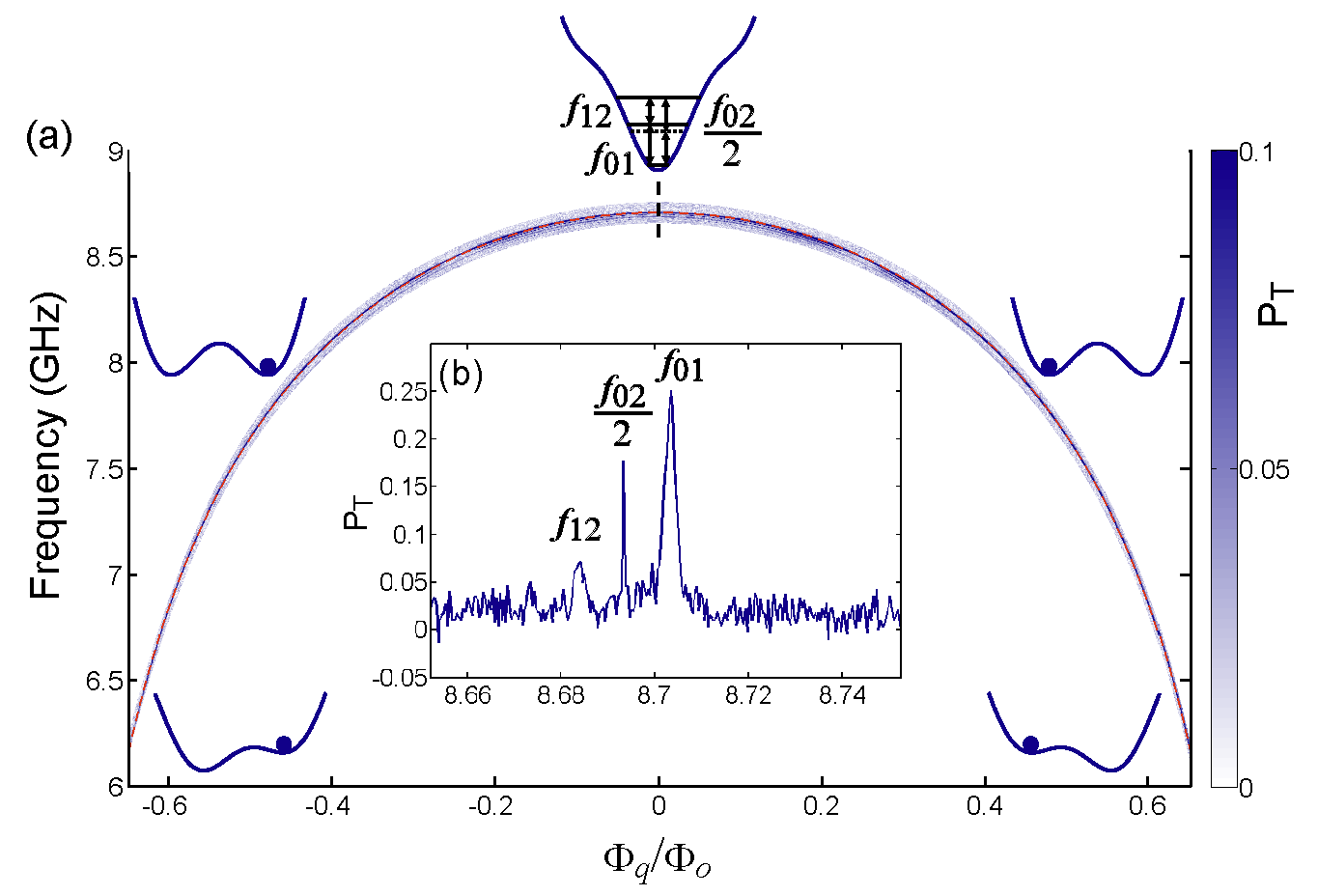}
  \caption{
(Color online) (a) Qubit spectroscopy (design $A$). The dashed line is a fit to the model described in the main text. Potential well configurations at
various flux biases are sketched with a dot denoting the
metastable minimum used for qubit operation. (b) The inset shows that
at the deepest well with a minimum anharmonicity
($\alpha_r=-0.2\%$), the qubit is still sufficiently coherent for
the $f_{01}$ transition to be spectroscopically distinguishable
from the $f_{12}$ (and $f_{02}/2$) transition, allowing for qubit
operations. Tracking these spectroscopic peaks along the full
spectroscopy provides us with a measure of the qubit anharmonicity
as shown later in Fig.~\ref{Fig9}.
 }
 \label{Fig5}
\end{figure*}

\begin{figure*}[t!]
  \centering
  \includegraphics[width=6in]{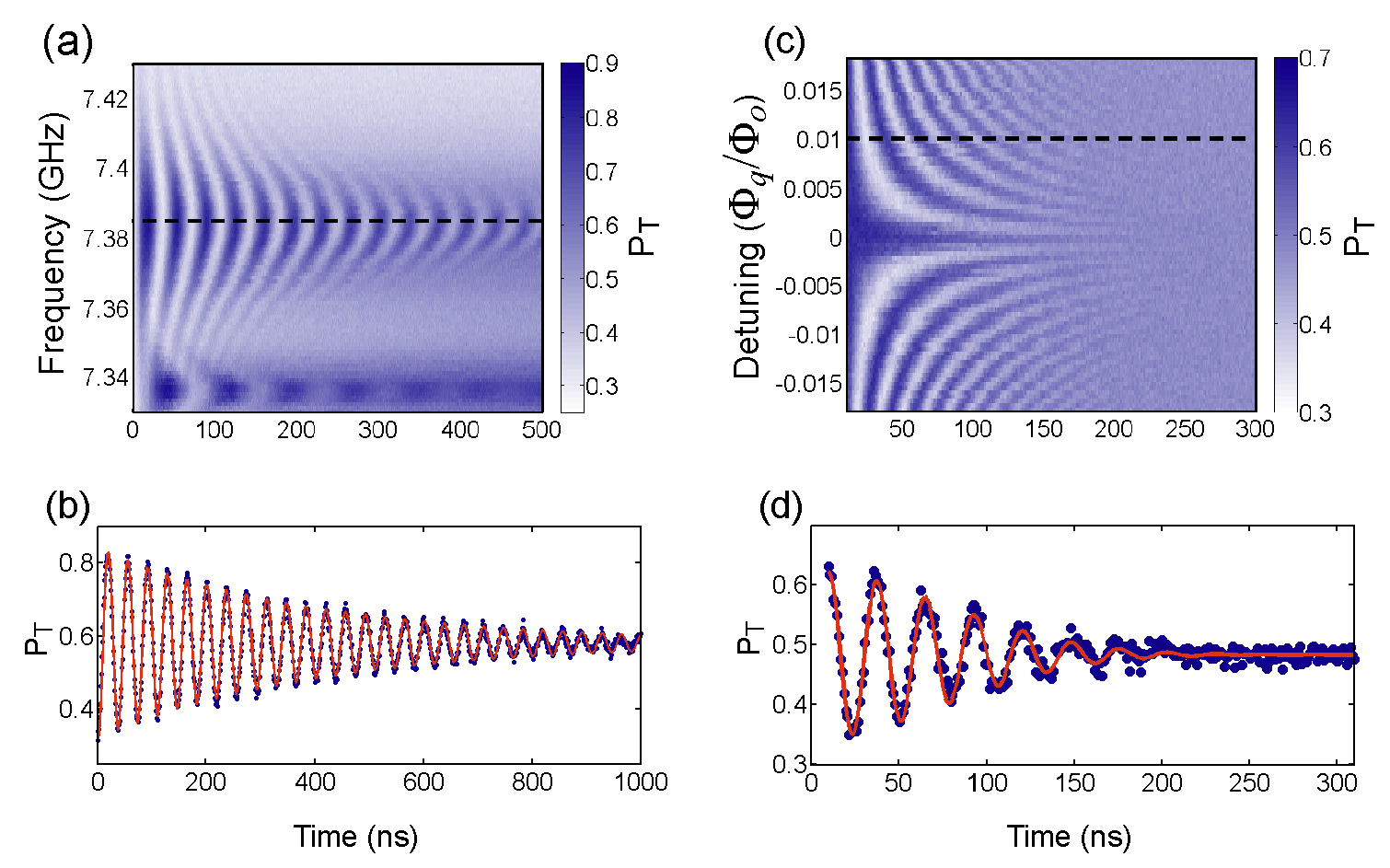}
  \caption{
(Color online) (a) Time domain measurements (design $A$). Rabi oscillations for frequencies near $f_{01}=7.38$ GHz. (b) Line-cut on-resonance along the dashed line in (a). The fit (solid line)
yields a Rabi oscillation decay time of $T'=409$~ns. (c) Ramsey
oscillations versus qubit flux detuning near $f_{01}=7.38$~GHz. (d) Line-cut along the dashed line in (c). The fit (solid line) yields a Ramsey decay
time of $T_2^*=106$~ns. With $T_1=600$~ns, this implies
\cite{WellstoodDecoherence} a phase coherence time $T_2=310$~ns.}
 \label{Fig6}
\end{figure*}

\begin{figure*}[t!]
  \includegraphics[width=6in]{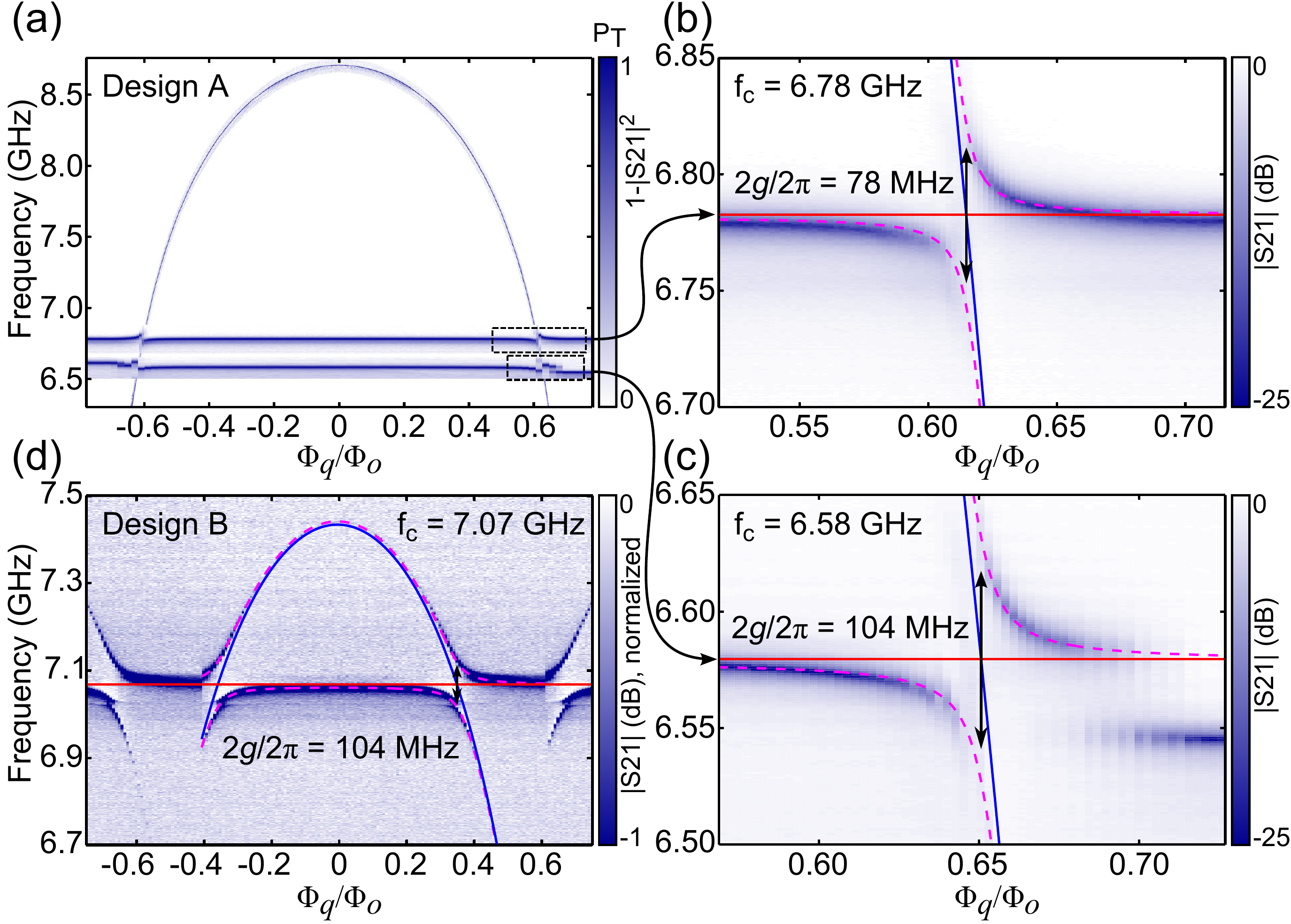}\\
  \caption{
(Color online) (a) Qubit spectroscopy (design $A$) overlaid with cavity spectroscopy at two frequencies, $f_c = $~6.58~GHz and 6.78~GHz. (b) Zoom-in of the split cavity spectrum in (a) when $f_c=$~6.78~GHz with corresponding fit lines. (c) Zoom-in of the split cavity spectrum in (a) when $f_c=$~6.58~GHz with corresponding fit lines. (d) Cavity spectroscopy (design $B$) while sweeping the qubit flux with $f_c=$~7.07~GHz showing a large normal-mode splitting when the qubit is resonant with the cavity. All solid lines represent the uncoupled qubit and cavity frequencies and the dashed lines show the new coupled normal-mode frequencies. Notice in (d) the additional weak splitting from a slot-mode just below the cavity, and in (c) and (d), qubit tunneling events are visible as abrupt changes in the cavity spectrum.}
 \label{Fig7}
\end{figure*}

The limiting factor for fast, dispersive qubit measurements has typically been a small $\kappa$, predominantly $\kappa/2\pi<5$~MHz. In practice, $\kappa$ is usually strategically reduced in order to avoid excessive loss from the Purcell effect, but as discussed above, this can not only reduce the speed of qubit measurements but also the SNR. In Ref.~\cite{PurcellFilter}, a ``Purcell filter'' allowed $\kappa/2\pi\approx 20$~MHz, with $T_1\approx 1\,\mu$s over a qubit spectral range of $1$~GHz. In Ref.~\cite{Johnson2012}, with $\kappa/2\pi=9$~MHz, single-shot fidelity near 90\% with a digitization bin-width of just 10~ns was achieved for $n>10$ and $T_1=1.8\,\mu$s. Our tunable cavity-QED approach provides a way to increase $\kappa$, while at the same time avoiding Purcell loss, and is more flexible than the ``Purcell filter'' \cite{PurcellFilter}. In this way, one can maintain a large qubit $T_1$, increase the SNR, and make faster qubit measurements, especially with quantum-limited parametric amplifiers. For this work, we operated a design with $\kappa/2\pi=24$~MHz, and achieved a rapid cavity response time with $2/\kappa\approx10$~ns. Although we did not test our system with a quantum-limited amplifier, we did show experimentally that \emph{on-average} state information could be acquired on a time scale of 10~ns, as discussed below.

\section{EXPERIMENTAL REALIZATION}
\label{(CD)}

We implement the tunable-cavity QED architecture as shown in
Fig.~\ref{Fig1}. Both the phase qubit and the cavity are based
on an \mbox{rf SQUID} design that allows rapid, precise, independent flux
control of either resonant frequency, while also providing a
convenient means for shared inductive coupling. Circuit
fabrication was performed with simple, single-layer aluminum
planar components on sapphire, and aluminum-oxide angle-evaporated
Josephson junctions (See Appendix~\ref{AA}). We performed
measurements on two device designs, $A$ and $B$, and found
consistent results for multiple samples (see Fig.~\ref{Fig1}).
The tunable cavity was coupled to the microwave feed-line through
a single coupling capacitor $C_{x}$, leading to a dip in
transmission on resonance. Sample boxes were mounted inside
thermal and magnetic shielding and attached to the mixing chamber
of a dilution refrigerator operated at 40~mK. The first stage of
amplification was performed with a superconducting SQUID-amplifier
also mounted at 40~mK with gain of $\sim$17~dB and noise
temperature near 1~K \cite{SQUIDAmp}. This was followed by a
HEMT-amplifier mounted at 4~K with a roughly 30~dB gain and then
further amplification at room temperature.

\subsection{Cavity Characterization}
\label{CC}
The tunable plasma frequency $f_c=\omega_c/2\pi$ of the \mbox{rf SQUID} cavity (neglecting the self-capacitance $C_J$ of the Josephson junction) is approximately given by 
\begin{equation}\label{eq:omegac}
\omega_c(\phi_c) = \frac{\omega_{c0}\sqrt{1+\beta_c\cos\delta_c}}{\sqrt{1 + \left(1+\beta_c\cos\delta_c\right)(L_s/\mathcal{L}_c)}}
\end{equation}
where $\omega_{c0}=1/\sqrt{\mathcal{L}_c C_c}$, $\mathcal{L}_c = L_c+M$, $\beta_c = \mathcal{L}_c/L_{Jc}$, and $L_{Jc} = \Phi_o/2\pi I_{oc}$ is the (zero-flux) Josephson junction
inductance, $I_{oc}$ is the cavity junction critical current, with the phase difference across the cavity junction $\delta_c$  determined by flux quantization, $2\pi\phi_c = \beta_c\sin\delta_c + \delta_c$. The spectroscopic response of each tunable cavity was measured by
monitoring the microwave transmission of a probe tone using a
network analyzer over a large range of flux biases spanning several flux quanta $\Phi_o$.
Fig.~\ref{Fig3} shows the frequency response for design $A$ (a similar
response was seen for design $B$). The cavity frequency $\omega_c$ has a maximum and a minimum value and is periodic in the total magnetic flux $\phi_c = \Phi_c/\Phi_o$ within the \mbox{rf SQUID} loop. The frequency dependence of the tunable cavity is made ``flatter'' near the maximum frequency due to the additional inductance $L_s$ in series with the cavity shunt capacitor $C_c$. Including $L_s$ reduces the participation ratio
of the nonlinear Josephson junction, helping to linearize the
cavity as well as reduce the effects of any possible dissipation
associated with the junction itself. A fit to the
spectroscopic data in Fig.~\ref{Fig3} (that includes the junction capacitance) gives $\mathcal{L}_c = 750$~pH, $L_s\approx
1.79$~nH, $I_{oc}\approx 0.31\,\mu$A, $C_c\approx 0.25$~pF, and $C_J=
20$~fF in agreement with design values. Here, $\mathcal{L}_c$ and $C_J$
were fixed at values determined by geometry using Fast Henry \footnote{http://www.fastfieldsolvers.com/} and the junction area
respectively (assuming $50$~fF/$\mu$m$^2$ for our Josephson
junctions plus approximately $4$~fF stray capacitance). The flux
periodicity provides a convenient means for determining the
flux coupling of the cavity bias coil, $M_{cB}=1.7$~pH.
The cavity response at the maximum frequency is shown in
Fig.~\ref{Fig3}(b--c). A skewed Lorentzian fit gives a loaded quality
factor of $Q_c=284$, with an internal quality factor of
$Q_i=3444$, and an external quality factor of $Q_e=309$, showing
that the cavity is strongly coupled to its feed-line. The highest spectroscopically measured quality
factors or minimum spectroscopic full-widths at half-maximum
(FWHM) were obtained at the maximum frequency, where the frequency
is first-order insensitive to magnetic flux noise.

The cavity response reveals weak coupling to resonant slot-modes
on-chip \cite{YaleMultiPurcell}. The number and strength of these modes was reduced by creating many aluminum wire bonds to stitch together all the
individual sections of ground planes \cite{YaleMultiPurcell,Chen2014}. These sections were a result of the single layer design and the need for on-chip bias lines and coplanar waveguide microwave launches. One of these slot-modes is
clearly visible as a blurry mode-splitting in the spectroscopy for
design $A$ at a fixed frequency of approximately $6.75$~GHz (see
Fig.~\ref{Fig3}(b)). The (lower-Q) slot-modes can contribute to additional
reductions in the qubit $T_1$ through an enhancement of the
Purcell effect at the specific slot-mode frequencies \cite{YaleMultiPurcell}. This is most visible as sharp dips in the $T_1$ values near the slot mode frequencies, shown later in Fig.~\ref{Fig10}(b) in Sec.~\ref{TCQEDC}.

\subsection{Qubit Characterization}
\label{QB}
The \mbox{rf SQUID} phase qubit \cite{SimmondsCoherentTLSs} has a potential energy curve that looks
like a folded washboard potential. When $\mathcal{L}_q/L_{Jq}\approx 3$, there are regions in flux where the potential has a single minimum
(single-valued) and it has regions with two minima
(double-valued). The phase qubit transition frequency $f_{01}=\omega_{01}/2\pi$ is approximately equal to the tunable plasma frequency $\omega_q/2\pi$ of the \mbox{rf SQUID}. Neglecting the self-capacitance of the Josephson junction, this is given by, 
\begin{equation}\label{eq:omegaq}
\omega_q(\phi_q) = \omega_{q0} \sqrt{\frac{1+(\beta_q+\beta_x)\cos\delta_q}{1+\beta_x\cos\delta_q}}
\end{equation}
where $\omega_{q0}=1/\sqrt{\mathcal{L}_q C_q}$, $\mathcal{L}_q = L_q+M$, $\beta_q = \mathcal{L}_q/L_{Jq}$, $\beta_x = L_x/L_{Jq}$, and $L_{Jq} = \Phi_o/2\pi I_{oq}$ is the (zero-flux) Josephson junction
inductance, $I_{oq}$ is the qubit junction critical current, with the phase difference across the qubit junction $\delta_q$  determined by flux quantization, $2\pi\phi_q = (\beta_q+\beta_x)\sin\delta_q + \delta_q$. When $\phi_q = 0$, the potential
is single-valued, symmetric, and nearly harmonic, yielding the
highest qubit frequency (see Fig.~\ref{Fig5} below). At $\phi_q = \pm 0.5$, the center of the
``overlap region" for the two co-existing flux states in the loop, the potential is double-valued, symmetric,
and yields a qubit frequency roughly half-way between the minimum
and maximum frequency (see Fig.~\ref{Fig5} below). This double-well
configuration is the most stable for storing information about
in which minima the system resides, with clockwise or
counter-clockwise circulating currents. The nearly $\Phi_o$
flux-difference within the \mbox{rf SQUID} loop between these two current
states makes this the ideal ``readout spot" for determining whether
any tunneling events have occurred between the two adjacent minima.

\subsubsection{Tunneling measurements with microwave readout}
\label{QBA}
During qubit operation, the flux is first set to $\phi_q = 0$ and
the system stays there long enough to reside in ground-state $|g\rangle$ due to energy relaxation with a characteristic energy lifetime, $T_1$. The flux is then adjusted to whatever operating flux value is desired. Following
any quantum operations, a tunneling measurement is initiated by a
fast flux measurement pulse \cite{TLSDecoherence, Fast} applied to
the qubit flux bias line. This pulse lowers the potential barrier
separating the metastable energy well from its neighboring well,
just enough that an excited qubit (in the $|e\rangle$-state or
higher) will tunnel to the adjacent well, shown as curved
arrows in Fig.~\ref{Fig4}(a) (and the $|g\rangle$-state tunneling
probability will typically be $\lesssim 10\%$, see Fig.~\ref{Fig4}(c)).
Following the measure pulse, the flux is then tuned to $\phi_q =
\pm 0.5$, the left (right) readout spot, when operating from
the right (left).

The two possible flux values at the readout spot leads to two
possible frequencies for the tunable cavity coupled to the qubit
loop. Similar microwave readout schemes have been used
with other rf-SQUID phase qubits \cite{RFReadout,
MartinisMultiplexed,Patel2013}.
For our circuit design, the size of this frequency difference is
proportional to the slope $df_c/d\phi_c$ of the cavity
frequency versus flux curve at a particular cavity flux
$\phi_c=\Phi_c/\Phi_o$. The transmission of the cavity can be
measured with a network analyzer to resolve the qubit flux (or
circulating current) states. The periodicity of the \mbox{rf SQUID} phase
qubit can be observed by monitoring the cavity's resonance
frequency while sweeping the qubit flux. This allows us to observe
the single-valued and double-valued regions of the hysteretic
\mbox{rf SQUID}. In Fig.~\ref{Fig4}(a), we show the cavity response to such a flux
sweep for design $A$. Two data sets have been overlaid, for two
different qubit resets ($\phi_q=\pm2$) and sweep directions (to
the left or to the right), allowing the double-valued or
hysteretic regions to overlap. There is an overall drift in the
cavity frequency due to flux crosstalk between the qubit bias line
and the cavity's \mbox{rf SQUID} loop that was not compensated for here.
This helps to show how the frequency difference in the overlap
regions increases as the slope $df_c/d\phi_c$ increases.

The optimization of the microwave readout of tunneling events
relies on maximizing the difference in microwave transmission at
an optimal cavity flux and cavity frequency. We place the cavity drive
frequency on the lower frequency dip and maximize the drive power
to a level where the nonlinearity of the cavity enhances peak
discrimination \cite{RFReadout}, but not so far that the dip-size
is reduced significantly. We then maximize the signal-to-noise
ratio for state discrimination by finding a optimal cavity flux $\phi_{c\,\rm{op}}$ that
maximizes the product: (dip-size/dip-width)$\times
(df_c/d\phi_c)$, where the dip-size and dip-width depend on
flux and drive power. As seen in Fig.~\ref{Fig4}(b), for measurements on design $A$, the dips can be well
separated with 100\% discrimination between the two possible
readout flux states. For this device, we found a maximum raw measurement
contrast of nearly 70\% for discriminating between the $|g\rangle$
and $|e\rangle$ states. This reduction is due to difficulties with eliminating
re-trapping effects \cite{MeasErrors} (see Fig.~\ref{Fig4}(c)), behavior that becomes for prevalent with rf SQUID phase qubits as $T_1$ increases.

\subsubsection{Qubit Spectroscopy}
\label{QBB}
Qubit spectroscopy was acquired for both designs $A$ and $B$. Like
the cavity, the qubit frequency $f_{01}$ is periodic in the
magnetic flux $\phi_q$ penetrating the \mbox{rf SQUID} loop, with a
maximum and a minimum operating frequency. However, in this case,
the minimum operation frequency is determined by the shallowest
metastable potential well with a $|g\rangle$-state tunneling
probability of $\gtrsim50$\%, which depends on the length of time
the metastable potential well configuration is maintained during
quantum manipulations \cite{Devoret}. We reset the qubit into a
single metastable well (or to a single current branch or ``current
step'', see Fig.~\ref{Fig4}) and then perform spectroscopy across a region
from the left-most step edge to right-most step edge. During
spectroscopy measurements, we apply an offset flux keeping the
qubit at its double-well, ``readout spot''. An arbitrary waveform
generator (Tektronix 5014B) was used to apply fast flux pulses to
move the qubit to fixed flux locations, for reset and for scanning
many potential well configurations for qubit operation
(see Fig.~\ref{Fig4}). At these locations, a spectroscopic microwave tone
is applied to the qubit for various frequencies in order to excite
the qubit transitions, followed by a fast measurement flux pulse. As
described above, any tunneling events are stored at the readout
spot. Fig.~\ref{Fig5}(a) shows an example of qubit spectroscopy for design
$A$. A fit to the spectroscopic data  
(that includes the junction capacitance) gives $L_q+M=
2.5$~nH, $L_x= 272$~pH, $I_{oq}\approx 0.33\,\mu$A, $C_q\approx
0.39$~pF, and $C_J= 20$~fF in agreement with design values. Here,
$L_q+M$ and $C_J$ were held fixed at values determined by
geometry using Fast Henry and the junction area respectively. The flux periodicity provides a
convenient means for evaluating the resultant flux coupling of the
qubit bias coil, $M_{qB}=10.9$~pH. The qubit response at the
maximum frequency is shown in Fig.~\ref{Fig5}(b) and shows that at the
deepest well configuration, the qubit is still sufficiently
coherent for the $f_{01}$ transition to be spectroscopically
distinguishable from the two-photon transition $f_{02}/2$ and the
next higher qubit level transition $f_{12}$. This provides a
measurement of the minimum relative anharmonicity,
$\alpha_r=(f_{12}-f_{01})/f_{01}=-0.2\%$. Tracking these
spectroscopic peaks along the full spectroscopy provides us with a
measure of the qubit anharmonicity across the full spectroscopic
range, as shown later in Fig.~\ref{Fig9}(a) in Sec.~\ref{TCQEDB}. For this demonstration, during the spectroscopic drive tone, the tunable cavity was rapidly shifted (via a flux pulse through $\phi_c$) to its minimum frequency $f_{c\,\textrm{min}}\approx 4.8$~GHz, in order to place it sufficiently far below all the qubit transition frequencies, providing a ``clean'' spectroscopic portrait of the phase qubit. Design $A$ had no visible spectroscopic splittings indicative of spurious two-level systems, while design $B$ showed one over the spectroscopic range from roughly 5.5~GHz
to 7.5~GHz \cite{SimmondsCoherentTLSs, TLSDecoherence}. This low occurrence of defects results from the use of small Josephson junctions ($\ll 1\,\mu$m$^2$). 

\subsubsection{Time-domain measurements}
\label{QBC}

An example of Rabi and Ramsey oscillations acquired with
tunnneling measurements are shown in Fig.~\ref{Fig6} around the qubit
resonance frequency, $f_{01}\approx 7.38$~GHz, with the tunable
cavity pulsed to its minimum frequency, $f_{c\,\textrm{min}}\approx 4.8$~GHz. 
Notice in Fig.~\ref{Fig6}(a) that Rabi oscillations at a lower
frequency are also visible for the two-photon transition, $f_{02}/2<f_{01}$. Fig.~\ref{Fig6}(b) shows a line-cut on-resonance with 
the qubit frequency $f_{01}$, along the dashed line in Fig.~\ref{Fig6}(a).
The Rabi oscillation amplitude decays exponentially in time with a
time-constant of $T'=409$~ns, as determined from the fit (solid
line). For Ramsey oscillations (shown in Fig.~\ref{Fig6}(c)), we placed a qubit frequency detuning z-pulse, applied with a fast, square flux pulse through $\phi_q$, between two $\pi/2$ microwave pulses and varied the
amplitude of the z-pulse. A fast-fourier
transform of these oscillations (not shown) confirms that the Ramsey frequency
matches the detuned qubit frequency during the z-pulse. Fig.~\ref{Fig6}(d)
shows a line-cut at a particular detuning, along the dashed line
in Fig.~\ref{Fig6}(c). The solid line represents a fit with a gaussian
decay envelope $\exp[-(t/T_2^*)^2]$, yielding an
inhomogenous-broadened dephasing time $T_2^*=106$~ns. Although we
did not perform a spin-echo measurement, we can get some
information about the coherence time $T_2$ from the exponential
decay of the Rabi oscillations when driven on-resonance
\cite{WellstoodDecoherence}, namely $T'=
(1/2T_{1}+1/2T_{2})^{-1}$. A separate measurement of the energy
decay of the qubit at this flux location gave $T_1=600$~ns, so that
$T_{2}\approx 310$~ns, about three-times the
inhomogenous-broadened value, or $T_{2}\approx 3\times T_2^*$.

In general, \mbox{rf SQUID} phase qubits have lower $T_2^*$ (and $T_2$)
values than transmons, specifically at lower frequencies, where
$df_{01}/d\phi_q$ is large and therefore the qubit is quite
sensitive to bias fluctuations and 1/f flux noise
\cite{FluxNoise}. For example, 600~MHz higher in qubit frequency,
at $f_{01}=7.98$ GHz, Ramsey oscillations gave $T_2^*= 223$~ns. At
this location, the decay of on-resonance Rabi oscillations gave
$T'=727$~ns, a separate measurement of qubit energy decay after a
$\pi$-pulse gave $T_1=658$~ns, and so, $T_{2}\approx 812$~ns, or
$T_2\approx 3.6\times T_2^*$, a small, but noticeable improvement over the lower frequency results displayed Fig.~\ref{Fig6}. The current device designs suffer from their planar geometry, due to a
very large area enclosed by the non-gradiometric \mbox{rf SQUID} loop
(see Fig.~\ref{Fig1}). Future devices will require
some form of protection against flux
noise \cite{FluxNoise}, possibly gradiometric loops or replacing the large geometric inductors with a much smaller series array of Josephson junctions \cite{Masluk2012}.  

\subsection{Tunable-Cavity QED Measurements}
\label{TCQEDA}

We can explore the coupled qubit-cavity behavior described by Eq.~(\ref{eq:H}) by performing spectroscopic measurements on either the qubit or the cavity near the resonance condition, $\omega_{01}=\omega_c$. Fig.~\ref{Fig7}(a) shows qubit spectroscopy for design $A$ overlaid with cavity spectroscopy for two cavity frequencies, $f_c = $~6.58~GHz and 6.78~GHz. Fig.~\ref{Fig7}(d) shows cavity spectroscopy for design $B$ with the cavity at its maximum frequency of $f_{c\,\rm{max}}=7.07$~GHz while sweeping the qubit flux bias $\phi_q$. In both cases, when the qubit frequency $f_{01}$ is swept past the cavity resonance, the inductive coupling generates the expected spectroscopic normal-mode splitting. \footnote{The weak additional splitting just below the cavity in Fig.~\ref{Fig7}(d) is from a resonant slot-mode.} We can determine the coupling rate $2g/2\pi$ between the qubit and the cavity by extracting the splitting size as a function of cavity frequency $f_c$ from the measured spectra. Three examples of fits are shown in Fig.~\ref{Fig7}(b--d) with solid lines representing the bare qubit and cavity frequencies, whereas the dashed lines show the new coupled normal-mode frequencies. For design $A$ ($B$), at the maximum cavity frequency of 6.78~GHz (7.07~GHz), we found a minimum coupling rate of $2g_{\rm{min}}/2\pi=78$~MHz (104~MHz). Notice that the splitting size is clearer bigger in Fig.~\ref{Fig7}(c) than for Fig.~\ref{Fig7}(b) by about 25~MHz. The results for the coupling rate $2g/2\pi$ as a function of $\omega_c/2\pi$ for design $A$ were shown in Fig.~\ref{Fig2} in section~\ref{TCQED}. Also visible in Fig.~\ref{Fig7}(c--d) are periodic, discontinuous jumps in the cavity spectrum. These are indicative of qubit tunneling events between adjacent metastable energy potential minima, typical behavior for hysteretic \mbox{rf SQUID} phase qubits
\cite{TLSDecoherence,RFReadout,MartinisMultiplexed}. Moving away
from the maximum cavity frequency increases the flux sensitivity,
with the qubit tunneling events becoming more visible as steps. This behavior is clearly visible in Fig.~\ref{Fig7}(c) and was already shown in Fig.~\ref{Fig4} in Sec.~\ref{QBA} and, as discussed there, provides a convenient way to perform rapid microwave readout of traditional tunneling measurements \cite{RFReadout,MartinisMultiplexed}. Next, we describe dispersive measurements of the phase qubit for design $A$. These results agree with the tunneling measurements across the entire qubit spectrum.

\subsubsection{Dispersive measurements of a phase qubit}
\label{TCQEDB}

\begin{figure}[!t]
  \includegraphics[width=3.3in]{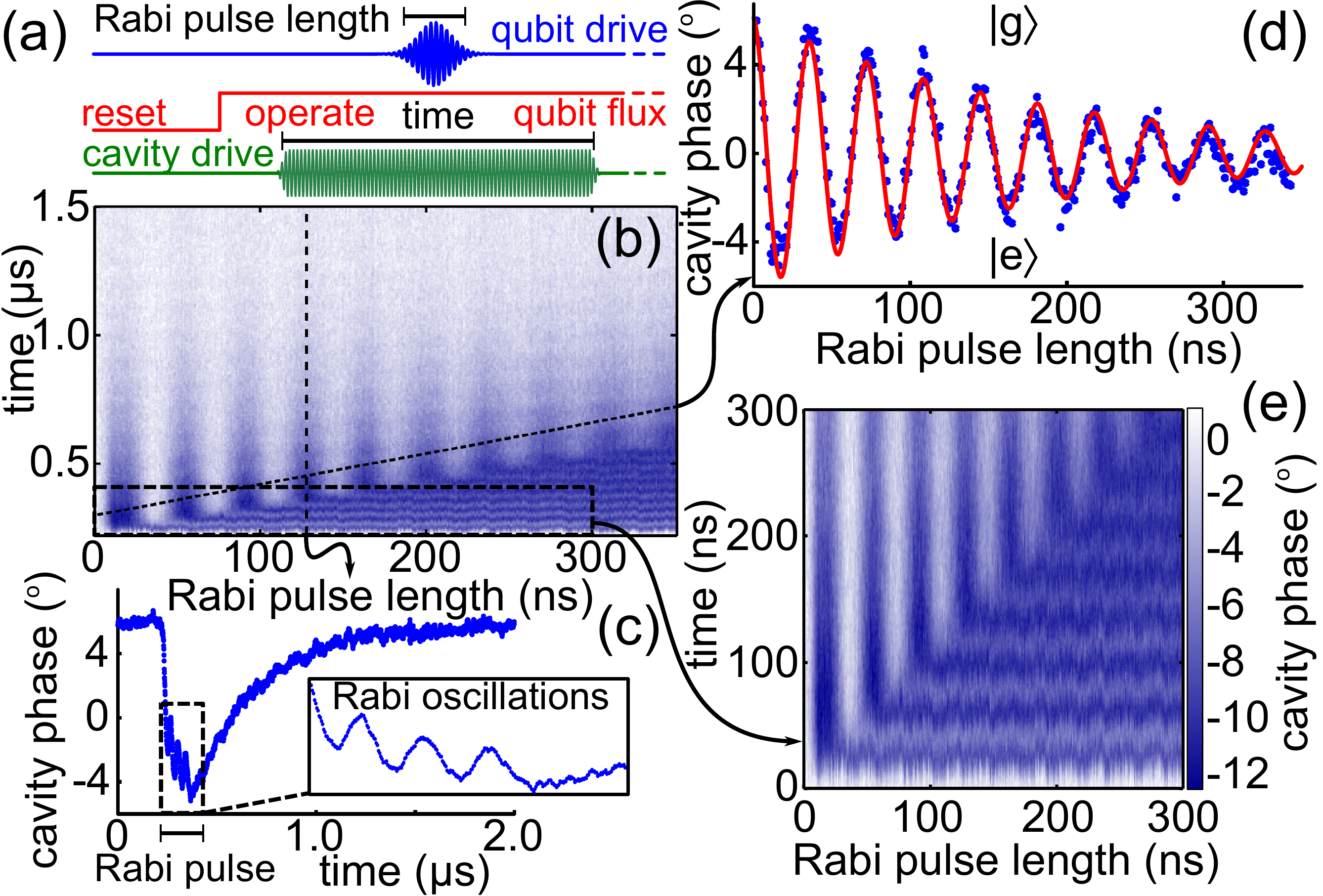}\\
  \caption{
(Color online) (a) Pulse sequence. (b) Rabi oscillations (design $A$) for various pulse durations obtained using dispersive measurement at $f_{01}=7.18$ GHz, with $\Delta_{01}=+10g$. (c) A single, averaged time trace along the vertical dashed line in (b). (d) Rabi oscillations extracted from
the final population at the end of the drive pulse, along the
dashed diagonal line in (b). (e) Zoom-in of dashed box in (b)
showing Rabi oscillations observed during continuous driving.}
 \label{Fig8}
\end{figure}

\begin{figure*}[!t]
  \includegraphics[width=5in]{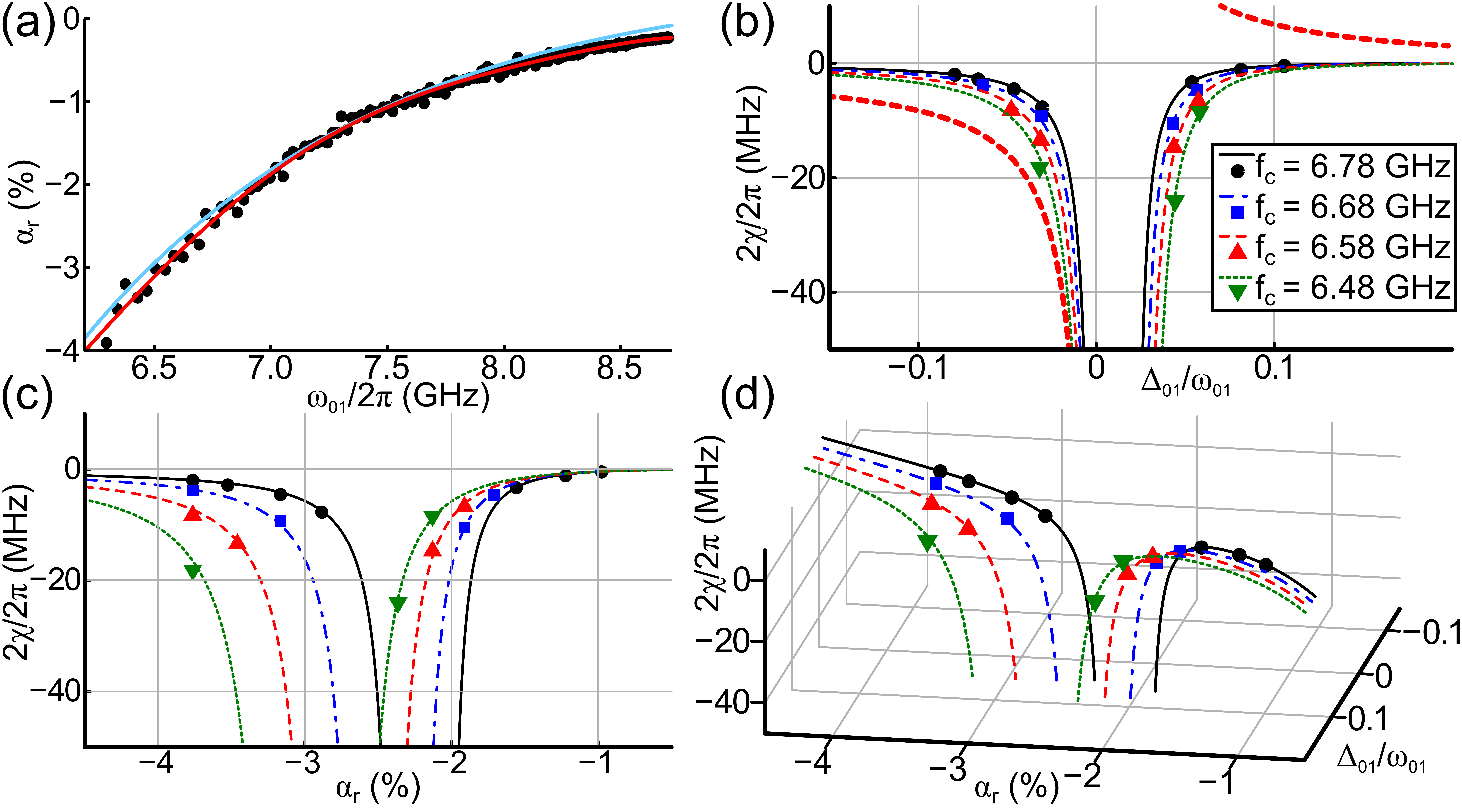}\\
  \caption{
(Color online) (a) Relative qubit anharmonicity $\alpha_r$ versus qubit frequency $\omega_{01}/2\pi$ (design $A$). The solid red line is a polynomial fit to the experimental data, used to calculate the three-level model curves in (b--d), while the blue line is a theoretical prediction of the relative anharmonicity (including $L_x$, but neglecting $C_J$) using perturbation theory and no fit parameters.
(b) Full dispersive shift $2\chi$ versus relative detuning $\Delta_{01}/\omega_{01}$ for four
different cavity frequencies, $f_c = $~6.78, 6.68, 6.58, and 6.48~GHz.
Symbols represent the data with lines showing the three-level
model predictions. The bold dashed line shows the two-level model
prediction when $f_c = 6.58$~GHz. (c) Full dispersive shift $2\chi$
versus the relative anharmonicity $\alpha_r$. (d) Full dispersive shift
$2\chi$ versus both the relative anharmonicity $\alpha_r$ and
relative detuning $\Delta_{01}/\omega_{01}$.}
 \label{Fig9}
\end{figure*}

We performed static dispersive measurements of the \mbox{rf SQUID} phase qubit during qubit manipulations by driving the cavity with a microwave tone and monitoring the phase response of the transmitted microwaves \cite{UnitVis}. The cavity frequency was completely fixed and driven near resonance, while the transmitted response was
sent to an $IQ$-mixer for homo-dyne detection with the quadrature
results captured with a digitizer card. The full qubit state-dependent
dispersive frequency shift $2\chi$ was inferred from a phase shift in the
outgoing microwave signal, $\phi = \pm\arctan(2\chi/\kappa)$, with
$\kappa = \omega_c/Q_c$ and $Q_c$ is the cavity's loaded quality
factor. Rabi oscillation data taken dispersively are shown in
Fig.~\ref{Fig8} with $f_{01}=7.18$~GHz and $\Delta_{01}=+10g$. The pulse sequence consists of a qubit reset,
followed by setting the operational qubit frequency, and then a
microwave (Rabi) drive is applied to the qubit for increasing
durations, while the cavity is monitored continuously. In
Fig.~\ref{Fig8}(b), we show the average phase response over
time for various durations of the Rabi pulse, with energy
relaxation after the drive. Strong coupling to
the cavity feedline ($\kappa/2\pi=24$~MHz) provides us with a fast cavity response time ($2/\kappa\approx 10$~ns), allowing us to capture time-averaged coherent Rabi oscillations during continuous microwave driving with evolution rates of approximately $\kappa$. This behavior can be seen clearly in the inset of Fig.~\ref{Fig8}(c), where we show a line-cut taken at a pulse duration
of 130~ns, and in Fig.~\ref{Fig8}(e) where we show a zoom-in of the oscillations seen in Fig.~\ref{Fig8}(b).  In Fig.~\ref{Fig8}(d) we show full amplitude Rabi oscillations, extracted from the data by taking a diagonal line cut, following the maximum displacement of the qubit state after the Rabi pulse, but before energy decay with $T_1 = 330$~ns. A fit gives a Rabi oscillation frequency
of 27.5 MHz, matching the time domain response during continuous driving
(shown in Fig.~\ref{Fig8}(e)), and an amplitude decaying with
$T'=180$~ns. This implies \cite{WellstoodDecoherence} a phase
coherence time $T_2=124$~ns, a value 1.5 times larger than the Ramsey
decay $T_2^*$ at this location, or $T_2\approx 1.5\times T_2^*$. As mentioned previously, coherence improves at higher qubit
frequencies and detunings.

Next, we carefully explore the size of the dispersive shifts for
various cavity and qubit frequencies. In order to capture the
maximum dispersive frequency shift experienced by the cavity, we
applied a $\pi$-pulse to the qubit. A fit to the phase response
curve \cite{UnitVis} allows us to extract the cavity's amplitude
response time $2/\kappa$, the qubit $T_1$, and the full dispersive
shift $2\chi$. Changing the cavity frequency modifies the coupling
$g$ and the detuning $\Delta_{01}$, while changes to the qubit
frequency change both $\Delta_{01}$ and the qubit's anharmonicity
$\alpha$. In Fig.~\ref{Fig9}(a), we show the phase qubit's anharmonicity as a function of its transition frequency $\omega_{01}/2\pi$ extracted from the spectroscopic data shown in Fig.~\ref{Fig5} from section~\ref{QBB} for design $A$. The solid red line is a polynomial fit to the experimental data, used to calculate the three-level model curves in Fig.~\ref{Fig9}(b--d), while the blue line is a theoretical prediction of the relative anharmonicity (including $L_x$, but neglecting $C_J$) using perturbation theory and the characteristic qubit parameters extracted section~\ref{QBB}. In Fig.~\ref{Fig9}(b--d), we find that the observed dispersive shifts strongly depend on all of these factors and agree well with the three-level model predictions \cite{Transmon, Boissonneault, ThreeLevelQubit}.
For comparison, in Fig.~\ref{Fig9}(b), we show the results for
the two-level system model (bold dashed line) when
$f_c=6.58$~GHz, which has a significantly larger
amplitude for all detunings (outside the ``straddling regime''). Notice that it is possible to increase the size of the dispersive shifts for a given $|\Delta_{01}|/\omega_{01}$ by decreasing the cavity frequency $f_c$, which increases the coupling rate $2g/2\pi$ (as seen in Fig.~\ref{Fig2} in section~\ref{TCQED}). Also, notice that decreasing the ratio of $|\Delta_{01}|/\omega_{01}$ also significantly increases the size of the dispersive shifts, even when the phase qubit's relative anharmonicity $\alpha_r$ decreases as $\omega_{01}$ increases. Essentially, the ability to reduce $|\Delta_{01}|$ helps to counteract any reductions in $\alpha_r$. These results clearly demonstrate the ability to tune the size of the dispersive shift through selecting the relative frequency of the qubit and the cavity. This tunability offers a new flexibility for optimizing dispersive readout of qubits in cavity QED architectures and provides a way for rf~SQUID phase qubits to avoid the destructive effects of tunneling-based measurements. 

\begin{figure*}[!t]
  \includegraphics[width=5in]{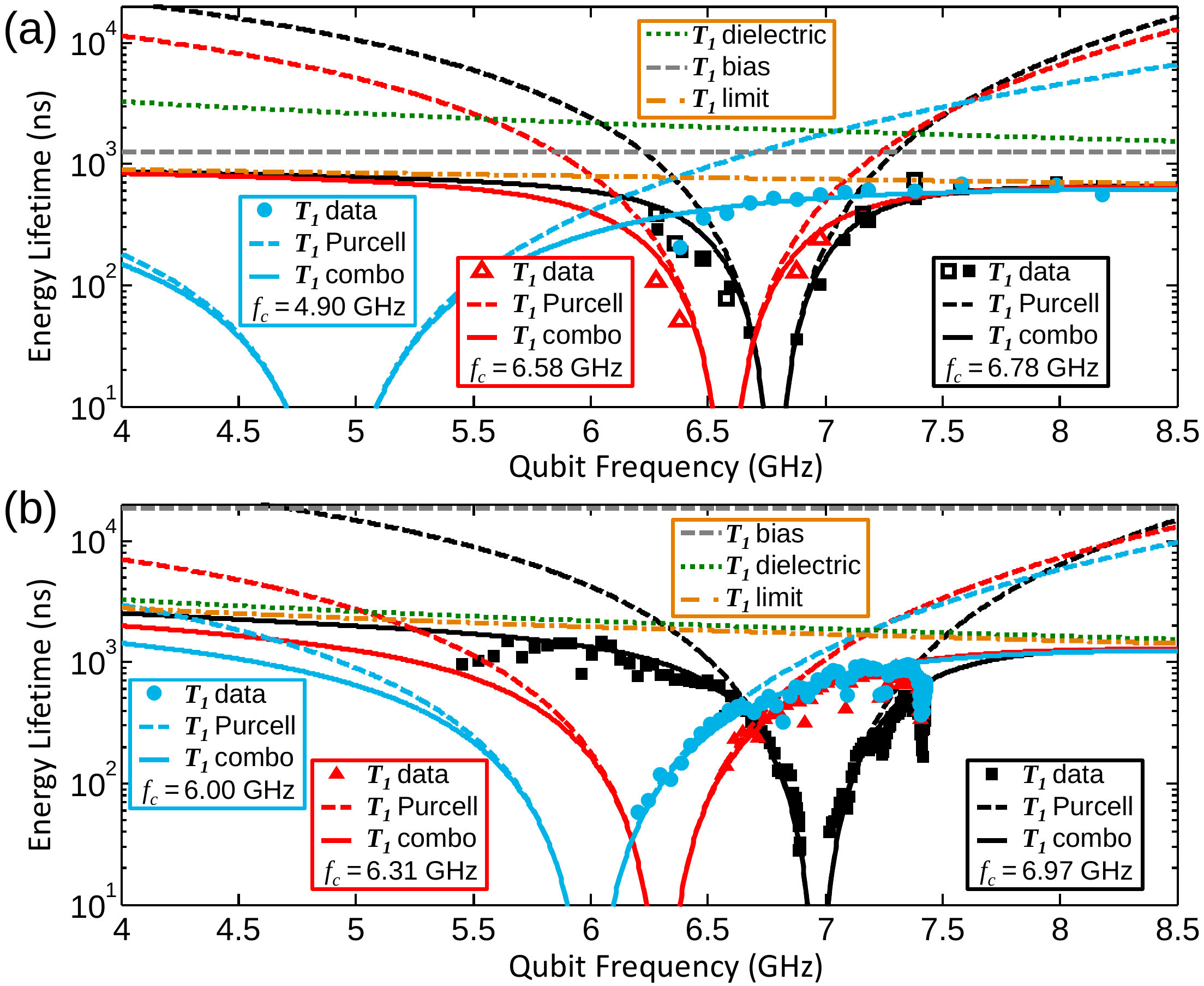}\\
  \caption{
(Color online) (a) The Purcell effect for design $A$. (b) The Purcell effect for design $B$. The solid (open) symbols represent data taken dynamically (statically) with tunneling (dispersive) measurement. The horizontal dashed
line shows the predicted decay time $T_1$ due solely to coupling
of the flux bias line to a 50$\,\Omega$ environment. The other dashed lines show $T_1$ from Purcell loss through the single-mode tunable cavity at each
frequency. The dotted lines are the predicted $T_1$ due to
dielectric loss with $Q_d=$~82,400. The dash-dot lines
represent the limiting $T_1$ due to coupling to the flux bias line
and dielectric loss. The solid lines represent the predicted $T_1$
by combining all the decay rates influencing the qubit: the flux
bias line, the dielectric loss, and the Purcell
effect (also see Table~\ref{tab:table1}).}
 \label{Fig10}
\end{figure*}

\subsubsection{Avoiding loss from the Purcell effect}
\label{TCQEDC}

It is possible to avoid loss from the Purcell effect through both static and dynamic operation of the tunable cavity. The experimental data we acquired for these two cases (described below), energy lifetime of the qubit ($T_1$) versus qubit frequency, looks identical, but the operational dynamics are obviously different. The ability to dynamically change the cavity frequency provides a way to isolate the qubit from the cavity at one time \emph{and} to optimize the cavity frequency to improve the quality of the qubit measurement at another time: \emph{both} avoiding Purcell loss during qubit operations and increasing the SNR for both tunneling and dispersive readout during qubit measurements. The qubit loss rate $\gamma_P$ via the Purcell effect \cite{YaleMultiPurcell} due to the single mode, tunable resonant frequency of the readout cavity increases as $g$ and $\kappa$ increase and decreases as $|\Delta_{01}|$ increases according to, 
\begin{equation}\label{eq:gammap}
\gamma_P=\frac{\left(g/\Delta_{01}\right)^2\kappa}{\left(1+\Delta_{01}/2\omega_c\right)^2}
\end{equation}
For our tunable-cavity QED system, this expression, along with Eq.~\ref{eq:g} for the tunable coupling strength $g$, determines how energy is lost by the phase qubit through the cavity for each cavity and qubit frequency. In order to avoid loss from the Purcell effect, the goal is to find each optimal cavity frequency for each possible qubit frequency, such that $\gamma_P$ is minimized. Generally, this is achieved by ensuring that there is a large frequency detuning $\Delta_{01}$ between the qubit and the cavity. However, for this system, the strong dependence of the qubit-cavity coupling $g$ on cavity frequency must be taken into account (see Fig.~\ref{Fig2}). Ideally, the qubit frequency should be placed far below the cavity frequency, so that $\Delta_{01}=\omega_{01}-\omega_c\ll 0$. However, it is clear, according to Eq.~\ref{eq:chi}, that the increase in $|\Delta_{01}|$ can also reduce the size of the dispersive shifts, reducing the SNR for dispersive readout. And as mentioned before, strong dispersive measurements go hand-in-hand with strong Purcell effects. Again, it is not possible, under static operation, to both avoid Purcell loss and maximize the strength of the dispersive qubit readout. And, as seen in section~\ref{QBA}, static operation, with varying cavity frequencies, can also reduce the effectiveness of the microwave readout of tunneling measurements. Thus, under static operations, we are forced to make a trade-off: longer coherence times or larger SNR. This balance may seriously reduce qubit coherence if one wishes to achieve single-shot dispersive readout \cite{Gambetta2008,Qjump}. The best one can hope to do is to minimize the Purcell losses to the point where other effects begin to dominate, while at the same time retaining a reasonable SNR. 

For our system, the total loss rate $\gamma_T = \gamma_{qB}+\gamma_{d}+\gamma_{P}$ is the summed combination of three contributions: (1) energy loss through coupling to the flux bias coil $\gamma_{qB}=(M_{qB}/\mathcal{L}_q)^2(1/Z_oC_q)$ (where $Z_o=50\,\Omega$),
(2) dielectric loss in the qubit $\gamma_{d}=\omega_{01}/Q_d$ (where
$\delta_d=1/Q_d$ is the effective dielectric loss tangent), and
(3) the Purcell loss rate $\gamma_P$. Because Purcell losses generally disappear rapidly as $|\Delta_{01}|$ increases, it is possible to find a minimum $|\Delta_{01}|$ where $T_1$ is mostly limited by other energy loss mechanisms. To characterize the energy loss in the qubit, we fully excite the qubit with a $\pi$-pulse and then measure the decay in time of the probability $P_T$ of finding the qubit in the excited state. Measurements were made over the entire qubit spectrum  for different cavity frequencies. Under static operation, the cavity was fixed at a set frequency $f_c$ throughout both qubit evolutions and measurement. Under dynamic operation, the cavity remained at the set frequency $f_c$ only during free-evolutions of the excited qubit, and would be rapidly flux-shifted to a new frequency optimized for qubit measurement. For these demonstrations, we performed dispersive readout only under static operation and tunneling readout under dynamic operation. This was convenient, as tunneling readout is fast and single-shot. Although tunneling measurements are ultimately destructive to the phase qubit, under dynamic operation we could still test our ability to both avoid Purcell loss, while still optimizing the readout conditions, as described in section~\ref{QBA}. In the future, with improved quantum-limited amplification, we plan to operate this system dynamically with dispersive readout.

\begin{table}[!t]
\caption{\label{tab:table1}%
Summary of Purcell results from fits of the predicted
``$T_1$ combo" values (including all the losses)
for both designs $A$ and $B$. For both design $A$ and $B$, the cavity frequencies were measured directly along with the coupling strengths $2g/2\pi$, except for design $A$'s lowest cavity frequency, when the qubit was unavailable due to tunneling of the \emph{metastable} ground-state $|g\rangle$. Here, the coupling strength was taken as a fit parameter. The ``$T_1$ bias'' due to the
qubit bias coil was calculated using the formula in the text
giving $1/\gamma_{qB}=1.25\,\mu$s ($18.5\,\mu$s) for design $A$
(design $B$). A single value of $Q_d=$~82,400 representing the
dielectric loss factor $\delta_d=1/Q_d$ was found to fit both data
sets well with ``$T_1$ dielectric" = $Q_d/\omega_{01}$.
}
\begin{ruledtabular}
\begin{tabular}{cccc}
 Circuit & $\omega_c/2\pi$ & $2g/2\pi$ & $\kappa/2\pi$ \\
 Design & (GHz) & (MHz) & (MHz) \\ 
\colrule
 A & 6.78 & 78 & 24\\ 
 A & 6.58 & 104 & 22\\ 
 A & 4.90 & 316 & 24\\
 \colrule
 B & 6.97 & 113 & 10\\ 
 B & 6.31 & 182 & 10\\ 
 B & 6.00 & 207 & 14\\ 
\end{tabular}
\end{ruledtabular}
\end{table}  

Data were acquired across the qubit spectrum for several well-separated cavity positions for both design geometries $A$ and $B$. The results agree well with
our model and are summarized in Table~\ref{tab:table1}. For design $A$, with the cavity placed at its maximum frequency $f_{c\,\rm{max}} = 6.78$~GHz and $\kappa/2\pi=24$
MHz, we find that the Purcell effect strongly reduces the combined
$T_1=1/\gamma_T$ over a significant portion of the qubit spectrum.
However, when $f_c = 4.9$~GHz, near the minimum cavity frequency, even with significantly stronger coupling $g/g_{\rm{min}}\approx 4$, qubit lifetimes are relatively large across the full qubit spectrum with a maximum value of
$T_1=0.72\,\mu$s, clearly limited by an over-coupled
flux bias line ($\gamma_{qB}$). For design $B$ (see Fig.~\ref{Fig1}(c)), we reduced the coupling to the
bias line by over a factor of 3 and lowered the maximum frequency of the qubit by over 1~GHz in order to take advantage of the inductive coupling, which improves operation when the qubit is mostly below the cavity. As seen in Fig.~\ref{Fig10}(b), we find significant improvement with a
maximum qubit lifetime of $T_1=1.5\,\mu$s, clearly limited by
dielectric losses with $Q_d=$ 82,400. This value, obtained for both design's, is consistent with losses due to two-level systems found in
single-layer aluminum lumped-element components of similar
dimensions (2 $\mu$m widths and gaps) \cite{DevoretGeometry}. Therefore, we
can estimate that \mbox{rf SQUID} phase qubits fabricated in a similar fashion, but with capacitor finger widths
and gaps $>10\,\mu$m should have increased qubit lifetimes
$>10\,\mu$s, as seen for planar lumped-element cavities and
transmon qubits \cite{DevoretGeometry, IBM,Pappas}. 

Thus, we have shown that it is possible to avoid loss from the Purcell effect, under both static and dynamic operation. It is important to note that: (1) for a given circuit design, Purcell losses can only be avoided when there exists a cavity position for each qubit position where Purcell loss does not dominate the qubit lifetime, and (2) the operational dynamics of the tunable cavity determine whether one can optimize qubit measurements along with reducing the effects of Purcell loss. These measurements show that the tunable cavity can be placed at its lowest or highest frequency in order to protect the qubit from Purcell losses over nearly the entire qubit spectrum with relatively large energy lifetimes. Energy loss is mostly limited by either strong flux bias coupling or dielectrics in the interdigitated capacitor, both of which can be improved through simple design changes. 

\section{SUMMARY AND CONCLUSIONS}
\label{SC}

In conclusion, we have developed a tunable-cavity QED architecture
with an improved \mbox{rf SQUID} phase qubit and a tunable lumped-element
cavity. Both tunneling and dispersive measurement techniques were
investigated. We have observed that qubit-cavity coupling $g$,
detuning $\Delta_{01}$, and qubit anharmonicity $\alpha$ all play
an important role in determining the size of the dispersive shifts
in this cQED system, as predicted. We have shown that by making the cavity
frequency rapidly tunable, it is possible to statically or dynamically tune both the qubit-cavity coupling and detuning to maximize qubit
performance during quantum evolutions, reducing unwanted Purcell effects
associated with the readout cavity. Dynamic operation of the tunable cavity with tunneling measurements has allowed us to both avoid Purcell loss and optimize the signal-noise-noise ratio for tunneling readout. The ability to avoid Purcell loss through the cavity relaxes design constraints on cavity-feedline coupling. This has allowed us to increase $\kappa>20$~MHz, significantly reducing the cavity's response time, increasing the maximum measurement bandwidth, critical for faster qubit measurements. Simple planar fabrication of the phase qubit leads to longer energy relaxations times, and although this represents only a modest improvement ($T_1\gtrsim 1\,\mu$s), the behavior of these phase qubits agrees well with the modeling of well understood dissipation mechanisms, in-line with the now ubiquitous transmon.

Unfortunately, the device designs is this work suffer from their planar, non-gradiometric geometry with very large enclosed areas ($\sim 200\times 300\,\mu$m$^2$), making them very susceptible to flux noise, whose dominating influence largely determines the phase coherence times. Thus, we were not able to directly verify the expected improvements to qubit dephasing times by reducing the effects of photon shot-noise from the readout cavity.  Obviously, because we tested only single qubit devices, we were also not able to directly verify the expected reductions in residual qubit-qubit bus coupling through control over the cavity's frequency. However, all the drawbacks associated with the circuit QED approach depend, for the most part, on the ratio of the coupling strength to the qubit-cavity detuning, $|g/\Delta_{01}|$. So that, showing a clear reduction in the energy lost by the qubit due to the Purcell effect through decreasing $|g/\Delta_{01}|$, we can infer that the two other sources of decoherence must also be naturally reduced in tunable-cavity QED systems.  

In the future, with improved device designs and multiple qubits, we hope to test the tunable-cavity QED concept more fully. And, by incorporating a wide-band quantum-limited parametric amplifier into the microwave readout chain, we should able to perform fast, pulsed dispersive readout with dynamic control over the cavity's frequency, taking full advantage of the benefits available to this architecture. Further design improvements should also push rf~SQUID phase qubit lifetimes above $10\,\mu$s. This bodes well for future experiments that require highly coherent rf~SQUID phase qubits to explore other types of rich physics. Moreover, future tunable-cavity QED devices can be designed to take further advantage of the tunable qubit-cavity coupling and should allow for significantly larger detunings providing even more protection from unwanted cavity effects. This work should
help to reduce spectral crowding, increasing the operational
bandwidth of multi-qubit systems, while providing faster qubit measurements.

\section{ACKNOWLEDGMENTS}
We thank \mbox{M. Castellanos-Beltran}, \mbox{M. Defeo}, and
\mbox{D. Slichter} for comments on the manuscript. This work was
supported by NSA under Contract No. EAO140639, and the NIST
Quantum Information Program. This Article is a contribution by NIST
and not subject to U.S. copyright.

\vspace{0.5cm}
\appendix*
\section{DEVICE FABRICATION}
\label{AA}
The devices were fabricated on sapphire wafers in two photolithography steps.
First, an aluminum base-layer was deposited by electron-beam
evaporation, then all wiring was patterned with a chlorine-based
gas etch. We used optical lithography and lift-off resist to
pattern a Dolan-bridge \cite{Dolan} and lightly cleaned it with an
oxygen plasma. The wafer was then placed in a custom-designed
electron-beam evaporator with an automated deposition and
oxidation system. Following a light ion-mill clean with a beam
(accelerator) voltage of 300~V (950~V) in $140\,\mu$Torr of argon
for 50 s, two nominally identical $Al/AlO_x/Al$ Josephson
junctions (one for the cavity and one for the qubit) were double
angle-evaporated in place. After the first aluminum layer, we used
thermal oxidation at room temperature in 760~mTorr of pure oxygen
for 10 minutes to provide a critical current density of approximately $1\,\mu$A/$\mu$m$^2$ for junctions with area approximately equal to $0.32\,\mu$m$^2$, giving critical currents of about $0.32\,\mu$A
and $L_J\approx 1$~nH. No insulators were deposited on the wafer
at any time during fabrication.  

\bibliographystyle{apsrev}
\bibliography{TunableCQED}

\end{document}